\documentclass[10pt,a4paper]{article}
 \usepackage[margin=3cm]{geometry}
\usepackage{graphicx}    
 \usepackage{amsmath,amsfonts, amsthm, hyperref}
 \usepackage{color}
 \definecolor{darkgreen}{rgb}{0.05,0.6,0.1}
 \definecolor{darkblue}{rgb}{0.05,0.1,0.6}
 \usepackage{enumerate,subfigure,array}
\usepackage{bbm}
  \newcolumntype{g}{>{$}c<{$}}

 \newcommand\enquote[1]{``#1''}
 
 \newcommand\be{\begin{equation}}
 \newcommand\ee{\end{equation}}

 \newcommand\llim[2]{\underset{#1 \rightarrow #2}{\text{lim}}}
 \newcommand\cvg[2]{\underset{#1 \rightarrow #2}{\longrightarrow}}
 \newcommand\tail[2]{\underset{#1 \rightarrow #2}{\sim}}
 \newcommand\EE[1]{\mathbb{E}\left[ #1 \right]}
 \newcommand\EC[2]{\EE{#1 \big| #2}}
 \newcommand\PP[1]{\mathbb{P}\left( #1 \right)}
 \newcommand\PC[2]{\PP{#1 \big| #2}}
 
 \def\RR   {\mathbb{R}}
 \def\Ff   {\mathcal{F}}
 
 \def\dd   {\textup{d}}
 
 \def\volRS   {\sigma^\text{RS}}

\makeatletter
\let\Hy@linktoc\Hy@linktoc@none
\makeatother

\begin{document}

\title{Quadratic Hawkes processes for financial prices}
\author{Pierre Blanc, Jonathan Donier, Jean-Philippe~Bouchaud
}
\date{\today}

\maketitle

\begin{abstract}
We introduce and establish the main properties of QHawkes (``Quadratic'' Hawkes) models. QHawkes models generalize the Hawkes price models introduced in E. Bacry et al. (2014), by allowing all feedback effects in the jump intensity that are linear and quadratic in past returns. A non-parametric fit on NYSE stock data shows that the off-diagonal component of the quadratic kernel indeed has a structure that standard Hawkes models fail to reproduce. Our model exhibits two main properties, that we believe are crucial in the modelling and the understanding of the volatility process: first, the model is time-reversal asymmetric, similar to financial markets whose time evolution has a preferred direction. Second, it generates a multiplicative, fat-tailed volatility process, that we characterize
in detail in the case of exponentially decaying kernels, and which is linked to Pearson diffusions in the continuous limit. Several other interesting properties of QHawkes processes are discussed, in 
particular the fact that they can generate long memory without necessarily be at the critical point. Finally, we provide numerical simulations of our calibrated QHawkes model, which is indeed seen to 
reproduce, with only a small amount of quadratic non-linearity, the correct magnitude of fat-tails and time reversal asymmetry seen in empirical time series. 
\end{abstract}



\section{Introduction: fBMs, GARCHs and Hawkes}

The hunt for a ``perfect'' statistical model of financial markets is still going on. Since the primitive Brownian motion model first proposed by Bachelier, 
droves of more and more sophisticated mathematical frameworks have been devised to describe the salient stylized facts of financial time series, namely: 
fat (power-law) tails of the return distribution, volatility (or trading activity) clustering with slow decay of correlations, negative return-volatility correlations (the so-called leverage effect), etc. The two most successful family of models to date are: a) GARCH-like models with slowly decaying memory kernels (e.g. FIGARCH
models) and b) stochastic volatility models where the log-volatility follows a fractional Brownian motion with a small Hurst exponent (e.g. the
Multifractal Random Walk \cite{bacry2001modelling} or, more recently, the ``rough volatility'' model of Gatheral, Jaisson and Rosenbaum \cite{gatheral2014volatility}). Although these models are remarkably 
parsimonious and convincingly capture many features of financial time series, they are still unsatisfactory on several counts. First, the returns in these 
models are conditionally Gaussian and therefore never ``fat enough'', even with a fluctuating volatility. Non-Gaussian residuals (or jumps) must be introduced by hand to match empirical probability distributions. Second, these models are not derived from deeper assumptions on the underlying mechanisms giving rise to fat-tails and volatility clustering. 
The theorist dream would be to start from, e.g. agents with simple trading rules or behavioral biases, and find that upon aggregation, their collective actions
lead to a certain class of stochastic model. Many attempts in this direction have been documented, in particular agent-based models of markets, stylized 
population dynamics models, or ``Minority Games'' -- for reviews see e.g. \cite{challet2013minority,cristelli2011critical}. Still, it is fair to say that none of these proposals has yet been widely accepted 
as a convincing ``micro-based'' explanation of the stylized facts recalled above.

A further, less discussed, but in our eyes highly relevant stylized fact is related to the time-reversal (a)symmetry (TRS/TRA) of financial time series. As initially emphasized by Zumbach \cite{zumbach2001heterogeneous} (following earlier ideas \cite{pomeau1982symetrie,ramsey1988characterization,ramsey1996time}), financial time series are {\it not} statistically symmetrical when past and future are interchanged; see
\cite{zumbach2009time}. There are (at least) two distinct effects that break this symmetry: one is the leverage effect alluded to above: {\it past} returns $r$ affect (negatively) {\it future} volatilities $\sigma$, but not the other way round. This is an effect that breaks both TRS and the up-down symmetry $r \to -r$. There is another effect though, that {\it is} invariant under $r \to -r$, namely: past large scale realized volatilities are more correlated with future small scale realized volatilities than vice-versa \cite{zumbach2001heterogeneous}. A more transparent way to explain this rather abstract notion is as follows: take $r$ to be daily returns (say) and $\sigma$ to be an estimator of volatility based on (say) five minute returns. Then consider, as in \cite{chicheportiche2014fine}, the average $\langle r_t^2 \sigma_{t+\tau}^2 \rangle_t$ with $\tau > 0$, which measures the correlation between past daily volatilities with future five minutes volatilities. The Zumbach effect, rephrased and empirically confirmed in \cite{chicheportiche2014fine}, is that 
$\langle r_t^2 \sigma_{t+\tau}^2 \rangle_t > \langle r_{t+\tau}^2 \sigma_{t}^2 \rangle_t$. It is clear that this criterion is invariant under $r \to -r$, and is
thus unrelated to the leverage effect. Where does such an asymmetry come from and what are the models consistent with TRA?

Interestingly, all continuous time stochastic volatility models, from the famous CIR-Heston model \cite{cox1985intertemporal,heston1993closed} to the Multifractal Random Walk model alluded to above, obey TRS 
by construction, and therefore {\it cannot} account for the empirical TRA of financial time series. GARCH-like models, on the other hand, do lead to strong TRA \cite{zumbach2001heterogeneous}, in fact stronger than seen in data \cite{chicheportiche2014fine}. This is expected; GARCH models do encode a feedback from past to future: large past realized returns lead to large future volatilities. This self-exciting mechanism is actually very similar to the one underlying ``Hawkes processes'' (invented in the context of
earthquake statistics), that have attracted a considerable amount of interest recently (for recent reviews, see \cite{bacry2015hawkes,laub2015hawkes}). In a financial context, Hawkes processes can be seen as a mid-way between purely stochastic models and agent based models.  One postulates that the activity rate at time $t$, $\lambda_t$, depends on the history of the point process itself $N_{s < t}$ via the auto-regressive relation 
\begin{equation}
\lambda_t = \lambda_\infty + \int_{-\infty}^t \phi(t-s) \ \dd N_s,
\label{eqn:Hawkes_model}
\end{equation}
where $\lambda_\infty$ is a baseline intensity and $\phi$ is a non-negative, measurable function such that $||\phi||_1 = \int_0^\infty \dd s \phi(s) \leq 1$. Hawkes processes are called ``self-exciting'', because every jump $\dd N_s \neq 0$ increases the probability of future events for $t>s$ via the kernel $\phi$; this in turn leads to activity clustering with an enticing causal interpretation: each event is a new signal for the rest of the market, triggering more activity. When calibrated on financial data, two remarkable features of the Hawkes process are found \cite{bremaud2001hawkes,hardiman2013critical,hardiman2014branching,bacry2015hawkes}: its kernel $\phi(s)$ shows long-range (power-law) decay $s^{-1-\epsilon}$, and its L1 norm $||\phi||_1$ is very close to unity, meaning that the process is on the verge of becoming unstable (see however \cite{filimonov2015apparent}). This is quite interesting since this is precisely the regime where the corresponding continuous time limit for the squared volatility (identified here with the activity) is a fractional CIR-Heston process \cite{jaisson2015rough}, with local Hurst exponent $H = \epsilon - 1/2$. This seems to close the loop: since $\epsilon$ is empirically found to be close to $1/2$ \cite{hardiman2013critical}, one has at hand a ``micro-model'' (the Hawkes process) that generates on coarse-grained scales a rough volatility process, which generalizes the CIR-Heston model to account for a slow, multi-timescale decay of volatility. Unfortunately, the situation is not as rosy yet: first, the fractional CIR-Heston process has tails 
that are much too thin (exponentially decaying) to account for the empirical distribution of volatility. Jaisson and Rosenbaum \cite{jaisson2015rough} therefore suggest to interpret the Hawkes process as a model for the {\it log-}volatility, but this is not natural. Second, following our discussion on TRS above, the fractional CIR-Heston process (as on fact the normal 
CIR-Heston one) is strictly TRS, and therefore fails to capture the observed TRA of financial time series!

The long story above sets the stage for our contribution, which is in an attempt to address the above deficiencies of the Hawkes formalism -- when applied to financial time series -- and take a step closer to the ``perfect'' model alluded to in our opening sentence. We propose a generalized version of the Hawkes process (called the 
QHawkes below) that includes features of the QARCH model introduced by Sentana~\cite{sentana1995quadratic} and revisited in depth in \cite{chicheportiche2014fine}. The idea is that the self-exciting mechanism is not only from market activity onto  market activity but also from actual price changes onto market activity. To make our motivation clear, consider a sequence of price moves, all with the same amplitude $|r| := \psi$. One expects that local trends, i.e. a succession of price moves in the same direction (up or down), triggers more volatility than a succession of compensated price moves, even though the high-frequency activity -- the number of price moves -- is exactly the same. The extra term we need to include in our generalized Hawkes process, beyond being motivated by empirical data, will encode mathematically this effect. We will see how this modification not only generates the needed fat tails in the return distribution (coming from the fact that the log-activity will indeed appear as a natural variable),
but also accounts quantitatively for the TRA of returns, at least on the intraday time scales on which we calibrate the model. We will see that in a particular case, the continuous-time limit of our model boils down to a simple, tractable two-dimensional Pearson diffusion, which can then be used as a low-frequency proxy for the volatility process.

The outline of the paper is as follows: we first introduce our general model in Section \ref{sec:qhawkes}, and highlight some of its core properties. 
Section \ref{sec:zhawkes} introduces a particular sub-family of factorized QHawkes processes, which we call ZHawkes after Zumbach, since it captures the Zumbach mechanism for generating TRA discussed above. Section \ref{sec:intra_QARCH} works out the parallel with QARCH models, which we calibrate on intra-day US stock data using the methodology similar to~\cite{chicheportiche2014fine}, showing a non-zero off-diagonal structure.  
We show in Section \ref{ZHawkes-exp} that in the case of exponential kernels the process is Markovian, and we write the corresponding stochastic differential equation as well as its continuous counterpart. Finally, we show in 
Section \ref{TRA_ZHawkes}, using numerical simulations, that 
the order of magnitude of the TRA generated by our ZHawkes process matches data quite well, and produce a volatility process with the right amount of 
fat-tails. Section \ref{sec:conclusion} then concludes.

\section{The QHawkes model}\label{sec:qhawkes}

\subsection{A general model}

Similarly to Hawkes processes~\eqref{eqn:Hawkes_model}, the QHawkes (Quadratic Hawkes) process $\left ( P_t \right )_{t\geq 0}$ is a self-exciting point process, whose intensity $\lambda_t$ is dependent on the past realization
of the process itself. As the name suggests, we model the intensity of price changes as the most general self-exciting point process that is Quadratic in  $\dd P_{s < t}$:
\begin{equation}
\lambda_t = \lambda_\infty 
\ + \ \frac1{\psi} \int_{-\infty}^t L(t-s) \ \dd P_s
\ + \ \frac1{\psi^2} \int_{-\infty}^t \int_{-\infty}^t K(t-s,t-u) \ \dd P_s \ \dd P_u,
\label{eqn:QHawkes_model}
\end{equation}
where $P$ is the high-frequency price, which is a pure jump process with signed increments. More precisely, whenever an event occurs between $t$ and $t + \dd t$, with probability $\lambda_t \dd t$, the price jumps by an amount $\xi$, where $\xi$ is a random variable of zero mean and variance $\psi^2$. A simple case that we
will consider below is $\xi = \pm \psi$ with probability $\frac12,\frac12$, where $\psi$ can be seen as the tick size. In the above equation, 
$L: \RR^+ \rightarrow \RR$ is a ``leverage'' kernel, coupling linearly price changes to market activity and 
$K: \RR^+ \times \RR^+ \rightarrow \RR$ is a quadratic feedback kernel. $\lambda_\infty$ is again the baseline intensity of the process (in the absence of any
feedback). Note that the above equation can be seen as a systematic expansion of the intensity of price changes in powers of past price changes, truncated to
second order. One could of course generalize the model further by adding, e. g. a third order term as $\int_{-\infty}^t \int_{-\infty}^t \int_{-\infty}^t K_3(t-s,t-u,t-v) \dd P_s  \dd P_u \dd P_v$, etc., but we will not consider this path further in the following.

Although it is necessary to account for the leverage effect on daily time scales, we will find later that on intra-day scales, the kernel $L$ is not significant, so for many applications one can focus on the quadratic kernel only and write
\begin{equation}
\lambda_t = \lambda_\infty 
\ + \ \frac1{\psi^2} \int_{-\infty}^t \int_{-\infty}^t K(t-s,t-u) \ \dd P_s \ \dd P_u.
\label{eqn:QHawkes_model_no_leverage}
\end{equation}
It is easy to see that model \eqref{eqn:QHawkes_model_no_leverage} is a generalisation of the simple Hawkes process for prices introduced in \cite{bacry2013modelling}: when choosing unit price jumps $\dd P_t = \pm \psi \dd N_t$ where $\psi$ can be seen as the tick and discarding any off-diagonal quadratic effects (so that $K(t,s) = \phi(t) \delta_{t-s}$), we recover a Hawkes process of kernel $\phi(s)=K(s,s)$.\footnote{In fact, if the kernels $K$, $K_3$, etc. to arbitrary order are all diagonal, the model boils down to a Hawkes process with leverage, i.e. $\lambda_t = \lambda_\infty + \int_{-\infty}^t \phi(t-s) \dd N_s
+ \psi^{-1} \int_{-\infty}^t L(t-s) \dd P_s$, with adequately redefined kernels $\phi$ and $L$, such that $\phi(s) - |L(s)| + \lambda_\infty \geq 0$ to ensure
positivity of the intensity.}

It is well known that the linear Hawkes process \eqref{eqn:Hawkes_model} can be seen as a branching process, where each \enquote{immigrant} event from the exogenous intensity $\lambda_{\infty}$ gives birth to a number of \enquote{children} events distributed as a Poisson law of parameter $n_H=||\phi||_1$, where $||\phi||_1$ is the $L^1$ norm of the kernel $\phi$. Each of these children in turn gives birth to a second generation of children with the same probability law and so on. 
When $n_H <1$, each immigrant gives birth on average to $n_H/(1-n_H)< \infty$ descendants. 
$n_H$ can thus be seen as a measure of endogeneity of the process, since it corresponds to the fraction of events that are triggered internally, reaching zero in the case of simple Poisson process and one in the special case of Hawkes process without ancestors \cite{bremaud2001hawkes}. 
The intuition behind the QHawkes in terms of a branching process is very similar, except that now the rate of events also depends on the interaction between the pairs of events. We will consider a positive feedback $K(s,t) \geq 0$ such that two mother events with the same sign (i.e. two prices moves in the same direction) increase the probability of a new event to be triggered in the future (i.e. increase future volatility), whereas compensated events have inhibiting effects, in line with (and directly motivated by) the empirical observations of \cite{chicheportiche2014fine}, as emphasized in the introduction.

\subsection{A special case: the ZHawkes model}\label{sec:zhawkes}

Motivated by the discussion in the introduction and by the empirical (intraday) results presented below, we will specialize the QHawkes model to the case where there is no leverage ($L \equiv 0$) and the quadratic feedback kernel $K$ is of the form
\begin{equation}
K(t,s) = \phi(t) \delta_{t-s} + k(t)k(s),
\nonumber
\end{equation}
i.e. the sum of a diagonal Hawkes component and of a factorisable, rank one kernel. We assume that $\phi,k : \RR^+ \rightarrow \RR^+$ are two positive, measurable functions that satisfy 
\begin{equation}
||\phi||_1 \equiv  \int_0^{+\infty} \phi(u) \ \dd u <+\infty
\quad , \quad
||k^2||_1 \equiv  \int_0^{+\infty} k(u)^2 \ \dd u <+\infty.
\nonumber
\end{equation}
Equation~\eqref{eqn:QHawkes_model} becomes in that case
\begin{equation}
\lambda_t = \lambda_\infty + H_t + Z_t^2,
\label{eqn:ZHawkes_model}
\end{equation}
where
\begin{itemize}
\item The \enquote{Hawkes term} is given by
\begin{equation}
H_t: = \int_{-\infty}^t \phi(t-s) \ \dd N_s; \qquad  N_t - N_{t^-} := \frac{1}{\psi^2} (P_t - P_{t^-})^2
\nonumber
\end{equation}
\item The \enquote{Zumbach term} given by $Z_t^2$ where
\begin{equation}
Z_t = \frac1\psi \int_{-\infty}^t k(t-s) \ \dd P_s.
\nonumber
\end{equation}
\end{itemize} 

We call $Z_t$ the Zumbach term since it is directly inspired by the series of empirical observations made by G. Zumbach on the volatility process (\cite{zumbach2001heterogeneous},\cite{zumbach2009time}). Indeed, $Z_t$ is simply a moving average of the past returns (with positive un-normalized weights $k(\tau)$). 
Therefore, this term will indeed be such that a sequence of returns in the same direction triggers more future volatility than compensated returns, as empirically observed \cite{zumbach2009time}.\footnote{Although Zumbach describes this effect at the daily time scale, whereas we will here study intra-day time scales.}

Besides its empirical motivations, the factorization property of the ZHawkes kernel significantly reduces the risk of over-fitting, since we will be left with two
one-dimensional kernels instead of the two-dimensional kernel in Eq.~\ref{eqn:QHawkes_model}.
As we see below, this simplified setup still captures the main phenomenology of price volatility,
with in particular time-reversal asymmetry and fat tails, {\it even for short-ranged kernels}.

\subsection{Mathematical framework}

Let us start by specifying the mathematical definition of the objects present in Equation~\eqref{eqn:QHawkes_model}:
\begin{itemize}
\item $(P_t)_{t \in \RR}$ is a pure jump process of stochastic intensity $(\lambda_t)_{t \in \RR}$, with unpredictable i.i.d. jump sizes $\xi$ of common law
$p$ on $(\RR, \mathcal{B}(\RR))$. We assume that $\int_{\RR} \xi \ p(\dd \xi) = 0$ and $\int_{\RR} \xi^2 \ p(\dd \xi) = \psi^2 < +\infty$, i.e. that jumps are centered and have a finite variance. 
\item $\mathcal{F}_t = \sigma(P_s, s\leq t)$ is the natural filtration of $P$.
\item $m(\dd t,\dd \xi)$ is the Punctual Poisson Measure associated to $P$, such that for all $t \in \RR$ and $A \in \mathcal{B}(\RR)$,
\begin{equation}
\llim{h}{0} \ \frac1 h \ \EC{m([t,t+h[,A)}{\Ff_t} = \lambda_t \ p(A).
\nonumber
\end{equation}
\end{itemize}

\noindent The quadratic kernel $K: \RR^+ \times \RR^+ \rightarrow \RR$ is assumed to satisfy
\begin{itemize}
\item Symmetry: $\forall s,t \geq 0, \ K(t,s) = K(s,t)$,
\item Positivity: $\forall f \in L^2(\RR^+), \ \int_0^{+\infty}\int_0^{+\infty} K(t,s) f(t) f(s) \ \dd t \ \dd s  \ \geq 0$,
\item Non-explosion: $\int_0^{+\infty} |K(t,t)| \ \dd t \ < +\infty$.
\end{itemize} 
$K$ defines an integral operator $T_K: L^2(\RR^+) \rightarrow L^2(\RR^+)$ which maps $f \in L^2(\RR^+)$ to
$T_K f: t \mapsto \int_0^{+\infty} K(t,s) f(s) \ \dd s$. If $K$ is continuous, this operator is Hilbert-Schmidt and thus compact and
one has $K(t,t) \geq 0$ for all $t \geq 0$ (see~\cite{buescu2004positive}). 
We define the trace of $K$
\begin{equation}
\text{Tr}(K) = \int_0^{+\infty} K(t,t) \ \dd t \ < +\infty.
\nonumber
\end{equation}
The leverage kernel $L: \RR^+ \rightarrow \RR$ is assumed to be a measurable function. By analogy with QARCH models (see~\cite{chicheportiche2014fine})
it should be dominated by $K$ in some way to ensure the positivity of the intensity $\lambda_t$ (see footnote 1 above). Since the leverage kernel is found empirically negligible in the sequel, we leave this positivity condition for future research.

\subsection{Necessary condition for time stationarity}

In the case of linear Hawkes processes, it has been shown that stationarity is obtained as soon as the norm of the kernel verifies $\left \| \phi  \right \|_1<1$. Intuitively, this means that each event triggers on average less than one child event, so that the clusters generated by each ancestor eventually die out. If this condition is violated, the probability that an ancestor generates an infinite number of events is non-zero, which can result in a stationary process only in the case $\left \| \phi  \right \|_1=1$ and $\lambda_{\infty}=0$ studied in \cite{bremaud2001hawkes}, see also \cite{hardiman2013critical}. Because of the quadratic feedback, the QHawkes process cannot be interpreted as a simple branching process, making things somewhat trickier. The goal of this section is to find a necessary condition for (first order) time stationarity.

\noindent We define the jump process $(N_t)$ that has the same jump times as $(P_t)$, with $\Delta N_\tau = (\Delta P_\tau)^2/\psi^2$ ($=1$ iff $\Delta P_\tau = \pm \psi$) for any jump time $\tau$,
and re-write Equation~\eqref{eqn:QHawkes_model} as
\begin{equation}
\lambda_t = \lambda_\infty + \mathcal{L}_t + H_t + 2 M_t
\label{eqn:QHawkes_decomp}
\end{equation}
with the notations

\begin{equation}
\left \{
\begin{array}{ll}
\mathcal{L}_t = \frac1\psi \int_{-\infty}^t L(t-u) \ \dd P_u & \text{(leverage)}\\[2mm]
H_t = \int_{-\infty}^t K(t-u,t-u) \ \dd N_u & \text{(Hawkes/diagonal)}\\[2mm]
M_t = \frac1{\psi^2} \int_{-\infty}^t \Theta_{t,u} \ \dd P_u & \text{(off-diagonal)}
\nonumber
\end{array}
\right.\\[2mm]
\end{equation}

\noindent where $\Theta_{t,u} = \int_{-\infty}^{u-} K(t-u,t-r) \ \dd P_r$ is $(\Ff_u)_{u\leq t}$-adapted for $t$ fixed. Since $P$ is a martingale, 
one has $\EE{M_t} = 0$ and $\EE{\mathcal{L}_t} = 0$. Therefore,
\begin{align}
\EE{\lambda_t} &= \lambda_\infty + \frac1{\psi^2} \EE{\int_{\RR} \int_{-\infty}^t K(t-s,t-s) \ \xi^2 \ m(\dd s,\dd \xi)} \nonumber \\
&= \lambda_\infty + \EE{\int_{-\infty}^t K(t-s,t-s) \ \lambda_s \ \dd s}
\nonumber
\end{align}
by definition of the punctual Poisson measure $m(\dd s,\dd \xi)$. We obtain
\begin{equation}
\EE{\lambda_t} = \lambda_\infty + \int_{-\infty}^t K(t-s,t-s) \ \EE{\lambda_s} \dd s.
\nonumber
\end{equation}
A necessary condition for the process $(\lambda_t)_{t \in \RR}$ to be in a stationary state is that its expected value $\overline{\lambda}\equiv \EE{\lambda_t}$ is constant,
positive and finite. This yields
$
\overline{\lambda} = \lambda_\infty + \overline{\lambda} \text{Tr}(K),
$
thus if $\lambda_\infty>0$,
\begin{equation}
\overline{\lambda} \ = \ \frac{\lambda_\infty}{1-\text{Tr}(K)}.
\nonumber
\end{equation}
This leads to the necessary stationarity condition\footnote{In the case of linear Hawkes processes, this condition is also sufficient to obtain stationarity in the case $\text{Tr}(K) < 1$ (whereas the case $\text{Tr}(K) = 1$ is more subtle, see \cite{bremaud2001hawkes}).}
\begin{align}
&\lambda_\infty>0 \text{ and } \text{Tr}(K) < 1\\
\text{ or } &\lambda_\infty=0 \text{ and } \text{Tr}(K) = 1.
\end{align}

In the special case of the ZHawkes process, the endogeneity ratio is then given by:
\begin{equation}
\text{Tr}(K) = ||\phi||_1+||k^2||_1 \equiv n_H + n_Z,
\nonumber
\end{equation}
where $n_H$ is the standard \enquote{Hawkes norm} while $n_Z \equiv ||k^2||_1$ is the \enquote{Zumbach norm}.

The existence of a finite average intensity $\overline{\lambda}$ is of course necessary for the process to reach a stationary state. However, the existence of higher moments of the intensity require stronger conditions on $K(t,s)$, similarly to the QARCH case studied in \cite{chicheportiche2014fine}. In particular, the decay of the off-diagonal 
part of $K$ must be fast enough to ensure the existence of two-point and three-point correlations of the process (see below).

\subsection{Auto-correlation structure in the QHawkes model}

It is quite useful for such type of models to investigate the relation between the input kernels and the auto-correlation functions of the generated process. Indeed, since the latter is directly observable on the data, the underlying kernels can then obtained by inverting such relations.  
For linear Hawkes processes, one finds a Wiener-Hopf equation that relates the two-points correlation function to the 1-d kernel \cite{bacry2012non}.
In our case, one also needs to consider the three-points correlation function, which will lead to a set of closed relations.

\subsection{Exact set of equations}

We take the model with no leverage, $L \equiv 0$. Equation~\eqref{eqn:QHawkes_decomp} becomes (see notations above):
\begin{equation}
\lambda_t = \lambda_\infty + H_t + 2 M_t.
\nonumber
\end{equation}
We define for $\tau \neq 0$ and $\tau_1>0, \tau_2>0, \tau_1 \neq \tau_2$, the correlation functions
\begin{align}
\mathcal{C}(\tau) &\equiv \EE{\frac{\dd N_t}{\dd t}\frac{\dd N_{t-\tau}}{\dd t}} - {\overline \lambda}^2
\ = \
\EE{\lambda_t  \frac{\dd N_{t-\tau}}{\dd t}} - {\overline \lambda}^2,
\nonumber \\
\mathcal{D}(\tau_1,\tau_2) &\equiv \frac1{\psi^2} \ \EE{\frac{\dd N_t}{\dd t}  \frac{\dd P_{t-\tau_1}}{\dd t}  \frac{\dd P_{t-\tau_2}}{\dd t}}
\ = \
 \frac1{\psi^2} \ \EE{\lambda_t  \frac{\dd P_{t-\tau_1}}{\dd t}  \frac{\dd P_{t-\tau_2}}{\dd t}}.
\label{eq:correl_C}
\end{align}
$\mathcal{C}$ is then extended continuously at zero, as in~\cite{H71}. Let us note that by construction $\mathcal{C}$ is even and $\mathcal{D}$ is symmetric. 
One finds the following exact equations between the auto-correlation functions ($\mathcal{C}$, $\mathcal{D}$) and the kernel $K$ (cf. the derivation in Appendix \ref{sec:correl_exact}):
\begin{align}
\mathcal{C}(\tau) &= \kappa \overline{\lambda} K(\tau,\tau) + \int_0^{\infty}  \dd u \, K(u,u) \mathcal{C}(\tau + u) + 2 \int_{0^+}^\infty \dd u \, \int_{u^+}^\infty  \dd r  \, K(\tau+u,\tau+r) \mathcal{D}(u,r),\label{eq:correls_C}\\
\mathcal{D}(\tau_1,\tau_2) = &2 K(\tau_1,\tau_2) [\mathcal{C}(\tau_2-\tau_1)+{\overline{\lambda}}^2] + \int_{(\tau_2-\tau_1)^+}^{\tau_2} \dd u \, K(\tau_2-u,\tau_2-u) \mathcal{D}(u-\tau_2+\tau_1,u) 
\nonumber \\
&+ 2 \int_{(\tau_2-\tau_1)^+}^{+\infty} \dd u\, K(\tau_1,\tau_2+u) \mathcal{D}(\tau_2-\tau_1,\tau_2-\tau_1+u),
\label{eq:correls_D}
\end{align}
where $\kappa = \frac1{\psi^4} \int_{\RR} \xi^4 \ p(\dd \xi)$ is the fourth moment of price jumps ($\kappa=1$ for constant price jumps). As $\mathcal{C}(\tau)$ and $\mathcal{D}(\tau_1,\tau_2)$ are directly measurable on the data, one can in principle infer the kernel $K(s,t)$ by inverting the above equations.

\subsubsection{Asymptotic behaviour}

Whereas the above equations \eqref{eq:correls_C} and \eqref{eq:correls_D} are difficult to solve in general, one can investigate the joint tail behaviour as $\tau\to\infty$ when both the kernel 
and the auto-correlation functions have power law decays. Let us assume that:
\begin{equation}
\left \{
\begin{array}{lll}
K(\tau,\tau) &\tail{\tau}{\infty} c_0 \ \tau^{-1-\epsilon} & \text{(diagonal, $\epsilon > 0$)}\\
K(\tau v_1, \tau v_2) &\tail{\tau}{\infty} \tilde{K}(v_1,v_2) \ \tau^{-2\delta} & \text{(off-diagonal, $\delta > 1/2$)}\\
\mathcal{C}(\tau) &\tail{\tau}{\infty} c_1 \ \tau^{-\beta} & \text{(2-points AC)} \\
\mathcal{D}(\tau, \tau) &\tail{\tau}{\infty} c_2 \ \tau^{-\beta'} & \text{(3-points AC, diagonal)}\\
\mathcal{D}(\tau v_1, \tau v_2) &\tail{\tau}{\infty} \tilde{\mathcal{D}}(v_1,v_2) \ \tau^{-2\rho} & \text{(3-points AC, off-diagonal)}\\
\end{array}
\right.
\end{equation}
where $c_0, c_1, c_2$ are constants and $\tilde{K}(v_1,v_2), \tilde{\mathcal{D}}(v_1,v_2)$ are bounded functions of $(v_1,v_2)$. The constraint $\epsilon > 0$
comes from the fact that $||\phi||_1$ must be finite, whereas $\delta > 1/2$ insures that the second and third moments are finite as well. We will furthermore 
assume (for simplicity) that $\epsilon < 1$, which is the interesting case in practice, and that the asymptotic behaviour of $K(\tau_1,\tau_2) \sim \tau_1^{-1-\epsilon}$ is 
restricted to a narrow channel around the diagonal $|\tau_1 - \tau_2| \ll \tau_1, \tau_2$, beyond which the off-diagonal power-law takes over.  

The exponents $\beta$ and $\rho$ can then be related to $\delta$ and $\epsilon$ by plugging these ansatzs into Eqs. \eqref{eq:correls_C} and \eqref{eq:correls_D} and
carefully matching the asymptotic behaviours. One finds several possible phases for the auto-covariance structure:

\begin{enumerate}

\item In the {\it non critical} case $\text{Tr}(K)<1$, we find: 
\begin{align}
\delta> (3 + \epsilon)/4 &\Rightarrow \beta = 1 + \epsilon; \qquad \beta' = 1 + \epsilon; \qquad \rho = \delta,
\label{eq:fast_decay_phase} \\
(2 + \epsilon)/3 <  \delta< (3 + \epsilon)/4 &\Rightarrow \beta = 4\delta-2;  \qquad \beta' = 1 + \epsilon; \qquad \rho = \delta,
\label{eq:slow_decay_phase1} \\
2 <  \delta < (2 + \epsilon)/3  &\Rightarrow \beta = 4\delta-2;  \qquad \beta' = 3 \delta - 1; \qquad \rho = \delta,
\label{eq:slow_decay_phase2}
\end{align}
The interpretation of these three phases is straightforward. In the first phase~\eqref{eq:fast_decay_phase}, the tail of the auto-correlation functions directly comes from the tail of the diagonal part of $K$: direct effects then dominate quadratic feedback effects. In the last two phases~\eqref{eq:slow_decay_phase1},\eqref{eq:slow_decay_phase2} however, a more sophisticated phenomenon comes into play, as off-diagonal effects feedback in such a way that they generate correlations with slower decay than that of the diagonal part of the kernel itself. In these phases, there is a possibility that $\beta < 1$ (corresponds to a long memory process) provided $\frac12<\delta<\frac34$. This result is important as it means that QHawkes processes {\it need not be critical} (i.e. $\text{Tr}(K)=1$) to generate long memory, unlike standard, linear Hawkes processes \cite{bremaud2001hawkes, saichev2010generation, hardiman2013critical, hardiman2014branching}. 

\item In the {\it critical} case $\text{Tr}(K) \to 1$, $\lambda_\infty \to 0$, the situation is subtler, as in the standard Hawkes case where the relation
between $\beta$ and $\epsilon$ completely changes, and the condition $0 < \epsilon < 1/2$ must hold for the process to even exist  \cite{bremaud2001hawkes}. In the present case, a similar mechanism operates and leads to:
\begin{align}
\delta> 3/4 &\Rightarrow \beta = 1 - 2 \epsilon; \qquad \beta' = 1 - \epsilon; \qquad \rho = \delta,
 \\
2/3 <  \delta < 3/4 &\Rightarrow \beta = 4 \delta - 2 \epsilon - 2;  \qquad \beta' = 1 - \epsilon ; \qquad \rho = \delta.
\\
(1+ \epsilon)/2 <  \delta < 2/3 &\Rightarrow \beta = 4 \delta - 2 \epsilon - 2;  \qquad \beta' = 3 \delta - \epsilon - 1; \qquad \rho = \delta.
\end{align}
provided $0 < \epsilon < 1/2$ and  $\delta > (1+ \epsilon)/2$, otherwise the critical process does not exist or is trivial. So in this critical case, 
the process is always long-memory (i.e. $\beta < 1$), or ceases to exist, as for the linear Hawkes process. 

\end{enumerate}

\section{The intra-day QHawkes model}\label{sec:intra_QARCH}

\subsection{QHawkes as a limit of QARCH}

In this section we investigate the link between the QHawkes model given by~\eqref{eqn:QHawkes_model} and the discrete QARCH model introduced by Sentana in \cite{sentana1995quadratic}, and revisited in depth in \cite{chicheportiche2014fine}. This will give us a way to calibrate the QHawkes on discretely sampled price time series. 
For a fixed time step $\Delta>0$, we define for all $t \in \RR$:
\begin{itemize}
\item the price (or log-price) increment between time $t$ and time $t+\Delta$: $r_t^\Delta = P_{t+\Delta} - P_t$,
\item the volatility at time $t$: $\sigma_t^\Delta = \sqrt{\EC{{r_t^\Delta}^2}{\Ff_t}}$.
\end{itemize}
The QHawkes model appears as the limit (in some sense) when $\Delta \rightarrow 0^+$ of the QARCH model
\begin{equation}
{\sigma_t^\Delta}^2 = {\sigma_\infty^\Delta}^2
\ + \ \underset{\tau \geq 1}{\sum} L^\Delta(\tau) \ r_{t-\tau \Delta}^\Delta
\ + \ \underset{\tau,\tau' \geq 1}{\sum} K^\Delta(\tau,\tau') \ r_{t-\tau \Delta}^\Delta r_{t-\tau' \Delta}^\Delta,
\label{eqn:equiv_QARCH}
\end{equation}
where ${\sigma_\infty^\Delta}^2 = \psi^2 \lambda_\infty \Delta, L^\Delta(\tau) = L(\tau \Delta) \ \Delta$
and $K^\Delta(\tau,\tau') = K(\tau \Delta, \tau' \Delta) \ \Delta$.
Indeed, for $t \in \RR$,
\begin{align}
\EC{{r_t^\Delta}^2}{\Ff_t} &= \psi^2 \ \PC{P_{t+\Delta} - P_t \neq 0}{\Ff_t} \ + \ o(\Delta)
\nonumber \\
&= \psi^2 \ \lambda_t  \ \Delta \ + \ o(\Delta),
\nonumber
\end{align}
which implies the scaling:
\begin{equation}
\frac{{\sigma_t^\Delta}^2}\Delta \cvg{\Delta}{0^+} \psi^2 \ \lambda_t.
\nonumber
\end{equation}
Thanks to this link between the two models, it is possible to calibrate a QARCH model on intra-day 5 minutes bin returns, as in \cite{BlancReport,chicheportiche2014fine}, and obtain some qualitative and quantitative insight on the
structure of the underlying QHawkes model. Indeed, the direct calibration of the latter would be significantly harder -- more noisy and computationally more
demanding -- and certainly beyond the scope of the present paper.

\subsection{Intra-day calibration of a QARCH model}\label{section:QARCH_calib}

QARCH models have mainly been calibrated on daily data so far (\cite{sentana1995quadratic}, \cite{chicheportiche2014fine}). To give a starting point to our study of quadratic effects in
high-frequency volatility, we calibrate a discrete QARCH on intra-day five-minute returns.

\subsubsection{Dataset and notations}\label{section:dataset}

We consider the same dataset as in~\cite{Allez_intra}, which is composed of stock prices on intra-day five-minute bins. It includes 133 stocks of the New York Stock Exchange, that have been traded without interruption between 1 January 2000 and 31 December 2009. This yields 2499 trading days, with 78 five-minute bins per day. For each bin, the open, close, high and low prices ($O,C,H,L > 0$) are available. We consider the log-price process and define on each bin:

\begin{itemize}
\item The return $r = \ln(C/O)$. 
\item The Rogers-Satchell volatility $\volRS = \sqrt{\ln(H/O)\times\ln(H/C) + \ln(L/O)\times\ln(L/C)}$.
\end{itemize}

\subsubsection{Normalization procedure}

 To be able to consider that intra-day prices are (approximately) independent realizations of a stationary stochastic process, we need to normalize the data carefully. As a matter of fact, strong intra-day seasonalities may corrupt the calibration results. This can be avoided to some extent through a cross-sectional and historical normalization. We take advantage of our large dataset to compute a cross-sectional intra-day volatility pattern for each trading day and we normalize the returns by this pattern, which dampens the effect of collective shocks on a given day. On the other hand, we use the intra-day/overnight model volatility model of~\cite{blanc2014fine} to factor out daily feedback effects and focus on pure intra-day dynamics.
To fully explain our normalization protocol, we introduce the following notations:
\begin{itemize}
\item The 5-min bin index $1\leq b \leq 78$, the day index $1 \leq t \leq 2499$ and the stock index $1 \leq u \leq 133$.
\item The empirical averages: $\langle x_{u,t,.} \rangle$ means conditional average of $x$ over bins, for stock $u$ and day $t$ fixed ; 
$\langle x_{u,.,b} \rangle$ and $\langle x_{.,t,b} \rangle$ are defined similarly as the conditional averages over days/stocks;
$\langle x \rangle = \langle x_{.,.,.} \rangle$ means average of $x$ over stocks, days and bins.
\end{itemize}
We compute the cross-sectional volatility pattern of day $t$, that we use to normalize the data of stock $u$, as: 
\begin{equation}
b \in \{ 1,\cdots,78\} \ \mapsto \ v_{u,t}(b) \equiv  \sqrt{\langle r^2_{u'\neq u,t,b} \rangle}.
\nonumber
\end{equation}
For stock $u$, the value
$r^2_{u,t,b}$ is excluded from the average, so that the normalization protocol does not cap the large returns of stock $u$ artificially.
We also consider the open-to-close volatility $\sigma^\text{D}_{u,t}$ of day $t$ for stock $u$, as computed by the intra-day/overnight model of~\cite{blanc2014fine}
with the data of stock $u$ over the days $\{ 1,\cdots,t-1 \}$. For $t=1$, we fix $\sigma^\text{D}_{u,1} = 1$.

\noindent The normalization protocol is as follows: $\forall u,t,b$,
\begin{itemize}
\item $r_{u,t,b} \leftarrow r_{u,t,b}/\sigma^\text{D}_{u,t}, \quad \volRS_{u,t,b} \leftarrow \volRS_{u,t,b}/\sigma^\text{D}_{u,t}$, \hspace{0.5cm} (normalization by open-to-close volatility)
\item $r_{u,t,b} \leftarrow r_{u,t,b}/v_{u,t}(b), \quad \volRS_{u,t,b} \leftarrow \volRS_{u,t,b}/v_{u,t}(b)$. \hspace{0.5cm} (cross-sectional intra-day normalization)
\end{itemize}

\noindent We further exclude trading days that involve at least one bin where the absolute return is greater than the average plus six
standard deviations. This represents approximately $7\%$ of trading days, i.e. one day every three weeks. Combined with the cross-sectional pattern  normalization, this data treatment strongly dampens the impacts of exceptional news events, which would require a special treatment and that we do not aim to model here.
Eventually, we set the mean of the squares to one and the average return to zero to make the stock universe more homogeneous: $\forall u,t,b$,
\begin{itemize}
\item $r_{u,t,b} \leftarrow r_{u,t,b}/\sqrt{\langle r^2_{u,.,.} \rangle}$,
 so that $\langle r^2 \rangle = 1$,
\item $\volRS_{u,t,b} \leftarrow \volRS_{u,t,b}/\sqrt{\langle {\volRS}^2_{u,.,.} \rangle}$,
 so that $\langle {\volRS}^2 \rangle = 1$,
\item $r_{u,t,b} \leftarrow r_{u,t,b}-\langle r_{u,.,.} \rangle$
 so that $\langle r \rangle = 0$.
\end{itemize}

\subsubsection{Calibration results}\label{sec:calibration}

The calibration process is similar to \cite{chicheportiche2014fine} and \cite{blanc2014fine}. A first estimate of the kernels is obtained with the Generalized Method of Moments, which uses a set of correlation functions that are empirically observable. Then, using this estimate as a starting point, we use Maximum Likelihood Estimation, assuming that the residuals are t-distributed (which accounts for fat tails that remain in the residuals). This second step significantly improves the precision of the calibration results, compared to a solo GMM estimation.

\begin{figure}[!t]
		\center$
		\begin{array}{cr}
			\hspace{1.5cm}\includegraphics[width = 5.5cm, height = 5.5cm]{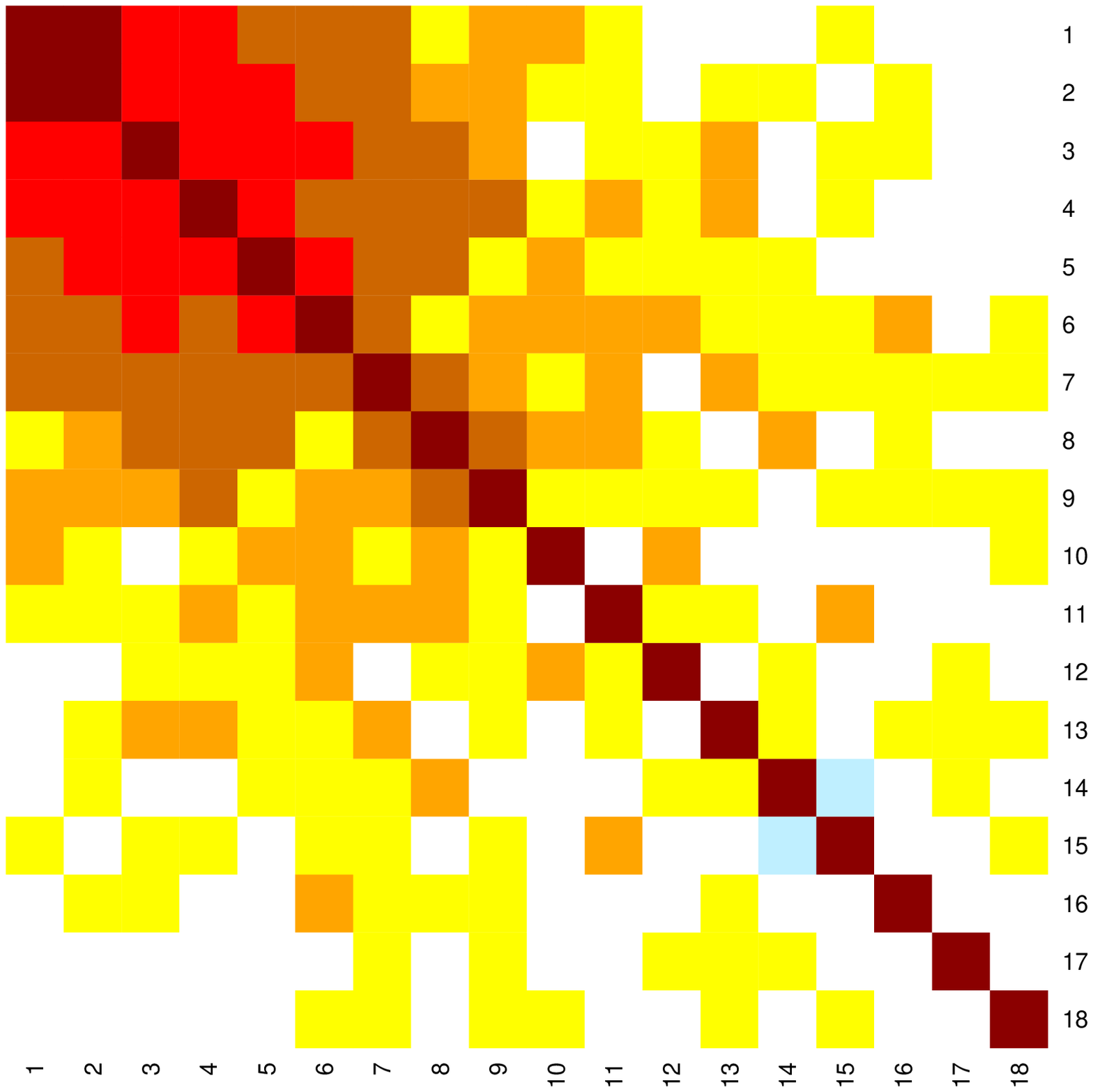}\hspace{.3cm}&
			\includegraphics[width = 0.49\textwidth]{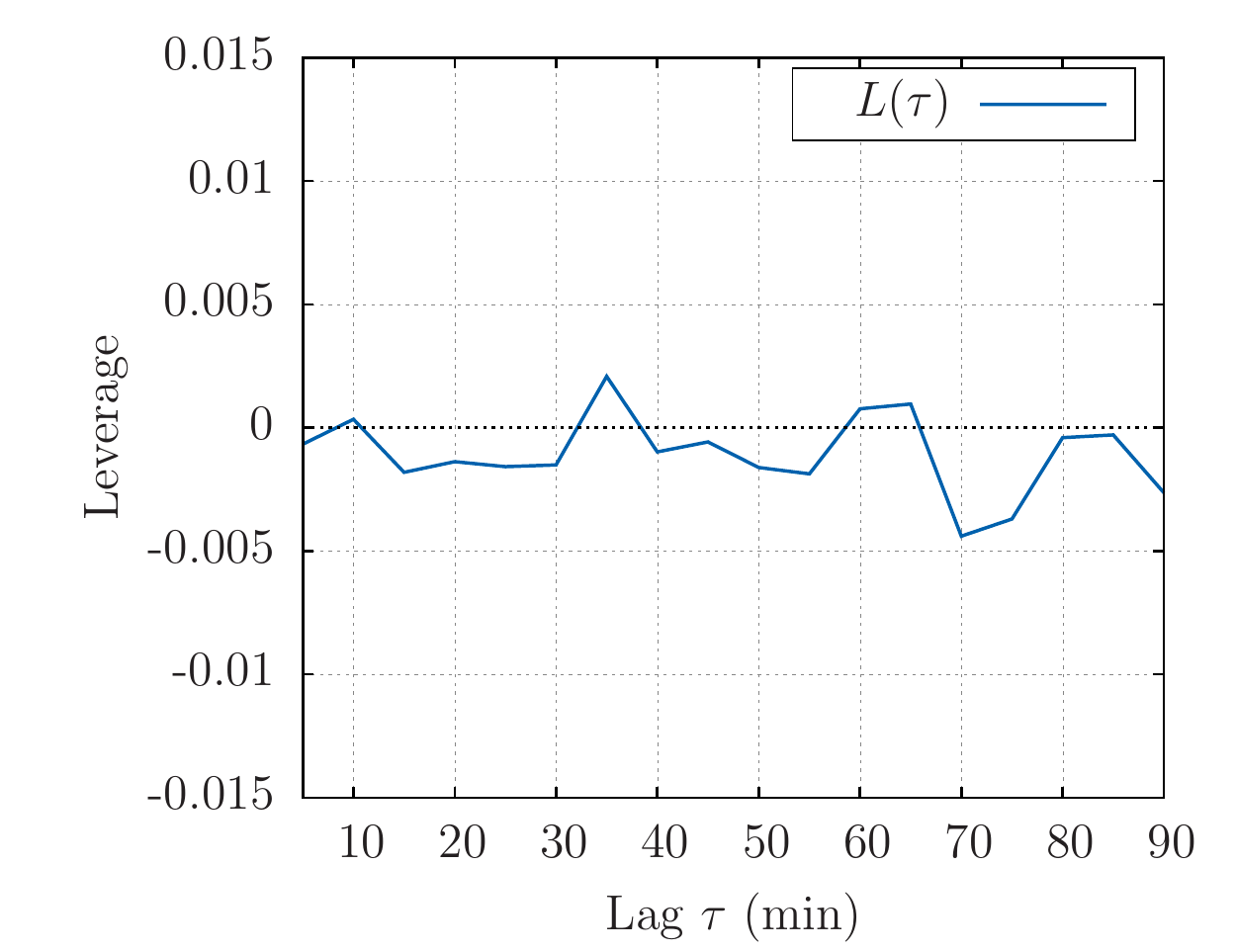}
		\end{array}$
		\caption{QARCH kernels calibrated on five-minute intra-day returns for US stocks.
		The maximum lag is $18$ bins, i.e. one hour and a half of trading time.
		Left: heatmap of the quadratic kernel. White coefficients are close to zero, blue ones are negative and yellow/orange/red ones are positive,
		with darker shades as they increase in absolute value.
		We see that all the significant coefficients are positive, with a non-negligible off-diagonal component.
		Right: leverage kernel. It is hardly distinct from zero and can be considered as pure noise (as opposed to daily models where it is significantly negative).}
        \label{graph:kernel_QARCH_US_stocks} 
\end{figure}

\begin{figure}[!t]
 \center$
		\begin{array}{cr}
			\hspace{1.5cm}
		\includegraphics[width = 5.5cm, height = 5.5cm]{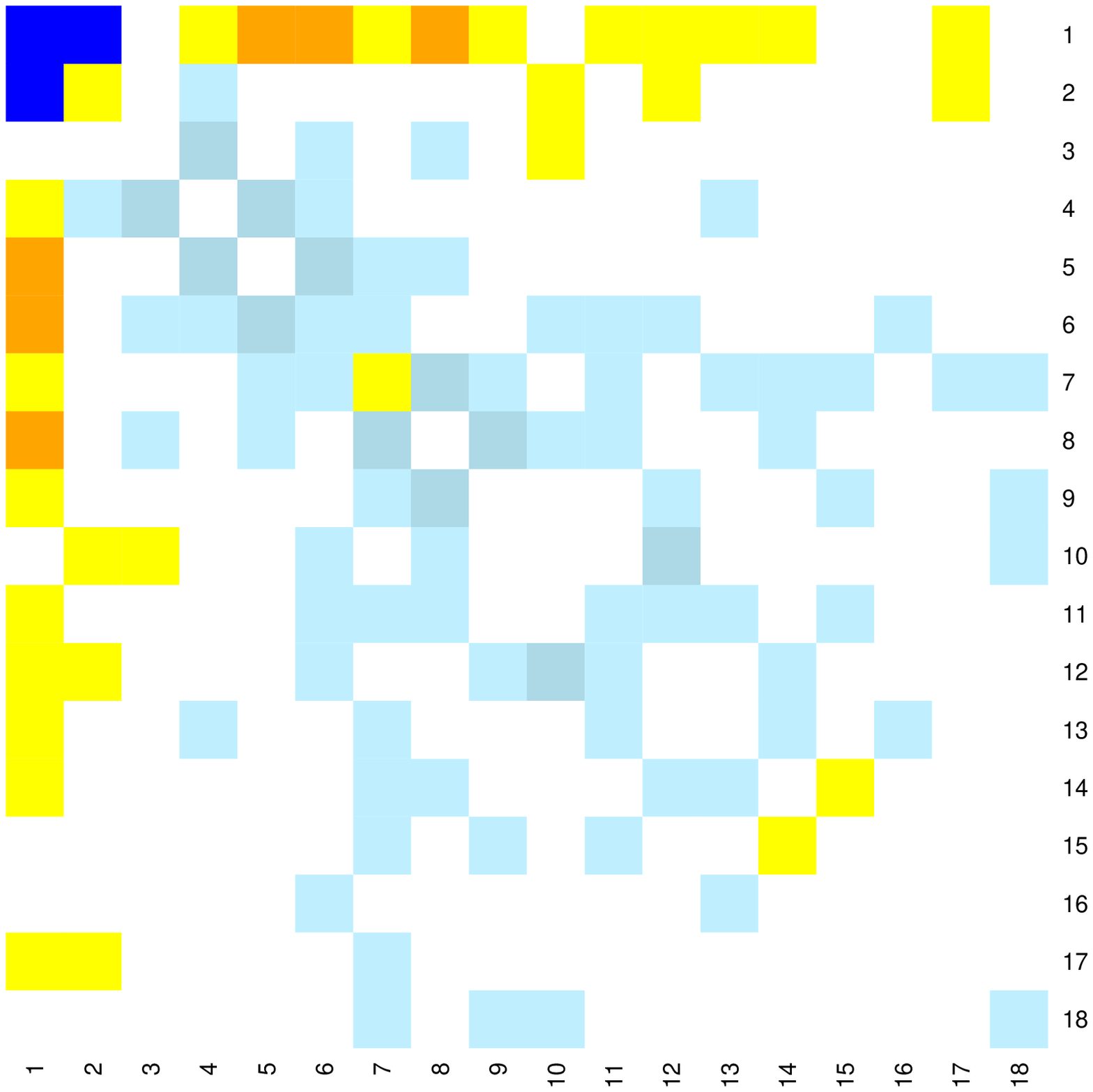}\hspace{.3cm}&
		\includegraphics[width = 0.49\textwidth]{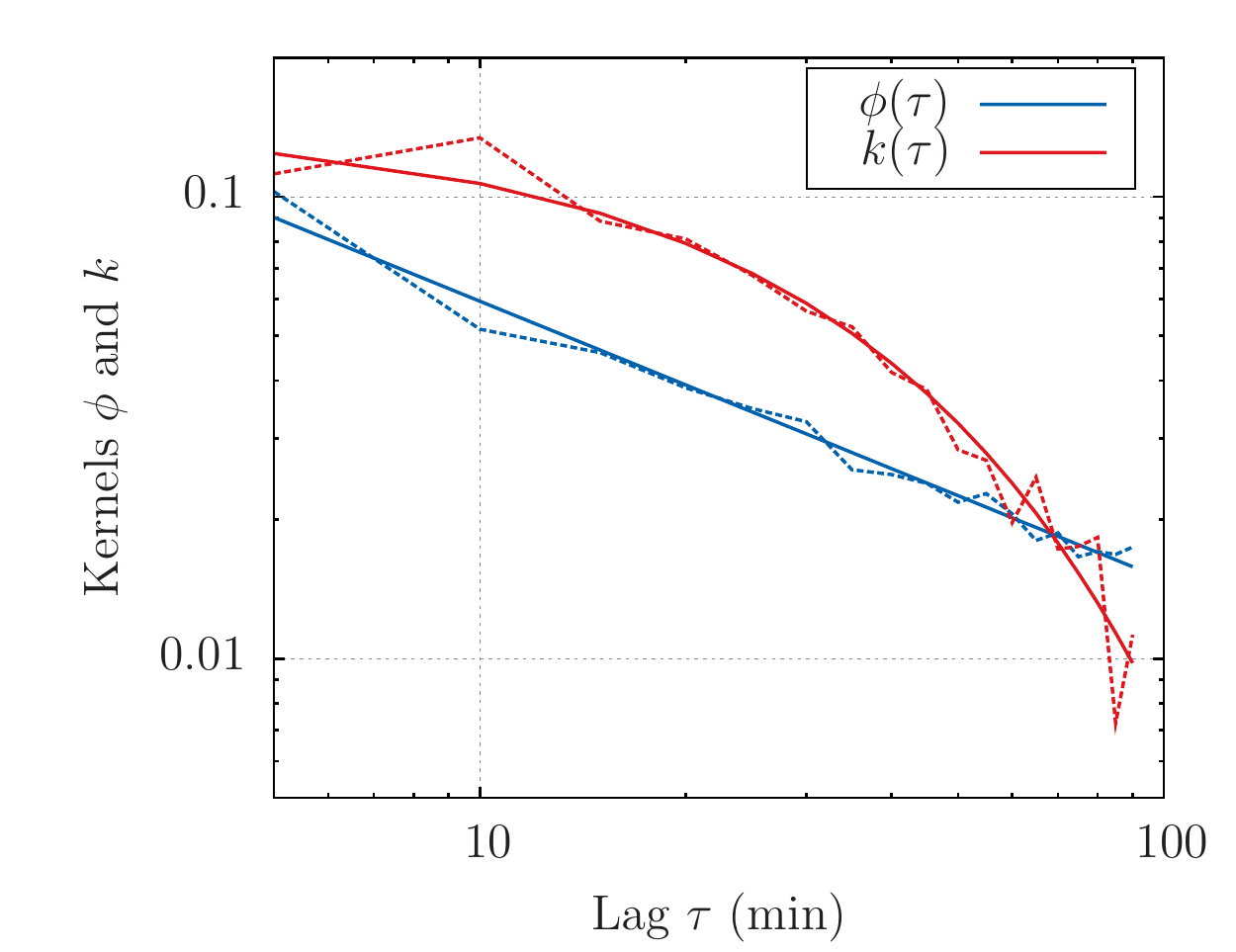}
		\end{array}$
		\caption{Fit of the kernel $K$ by the sum of a power-law diagonal matrix and an exponential rank-one matrix.
		Left: heatmap of the difference between the fitted matrix and the original one. The coefficients are small (white or lightly-colored)
		except for the upper-left corner: the original matrix features a stronger short-term feedback.
		Right: kernels $\phi(\tau)$ and $k(\tau)$ that minimize the matrix distance $\sum [K(\tau,\tau')-\phi(\tau) \delta_{\tau-\tau'} - k(\tau) k(\tau')]^2$. The rank-one kernel $k$ is plotted in red (and is larger for small $\tau$'s), and the diagonal kernel $\phi$ is plotted in blue, both in log-log scale.
		The dashed lines are the power-law fit for $\phi(\tau)$ with exponent $\alpha=0.6$, and the exponential fit for $k(\tau)$ with characteristic time about $30$ min.}
        \label{graph:ZHfit_QARCH_stocks} 
\end{figure}

We find it worth to notice that as opposed to the daily calibration
results of \cite{chicheportiche2014fine}, a clear off-diagonal structure appears in the feedback matrix in the intra-day case (see Figure~\ref{graph:kernel_QARCH_US_stocks}). Also, the intra-day leverage kernel is found to be close to zero, justifying the fact that we mainly consider $L\equiv 0$ throughout the paper.
\begin{figure}[!t]
 \center
		\includegraphics[width = 0.49\textwidth]{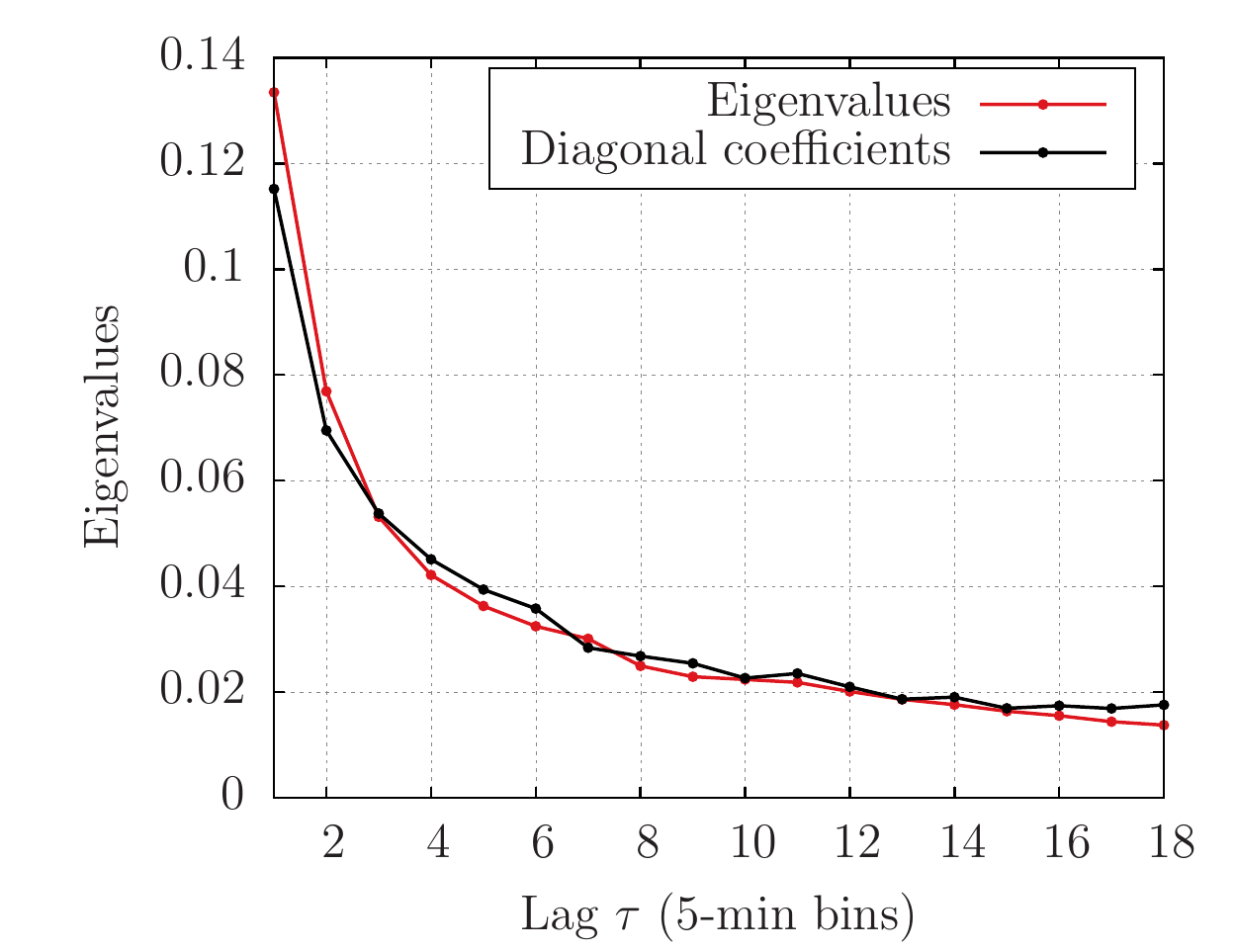}
		\includegraphics[width = 0.49\textwidth]{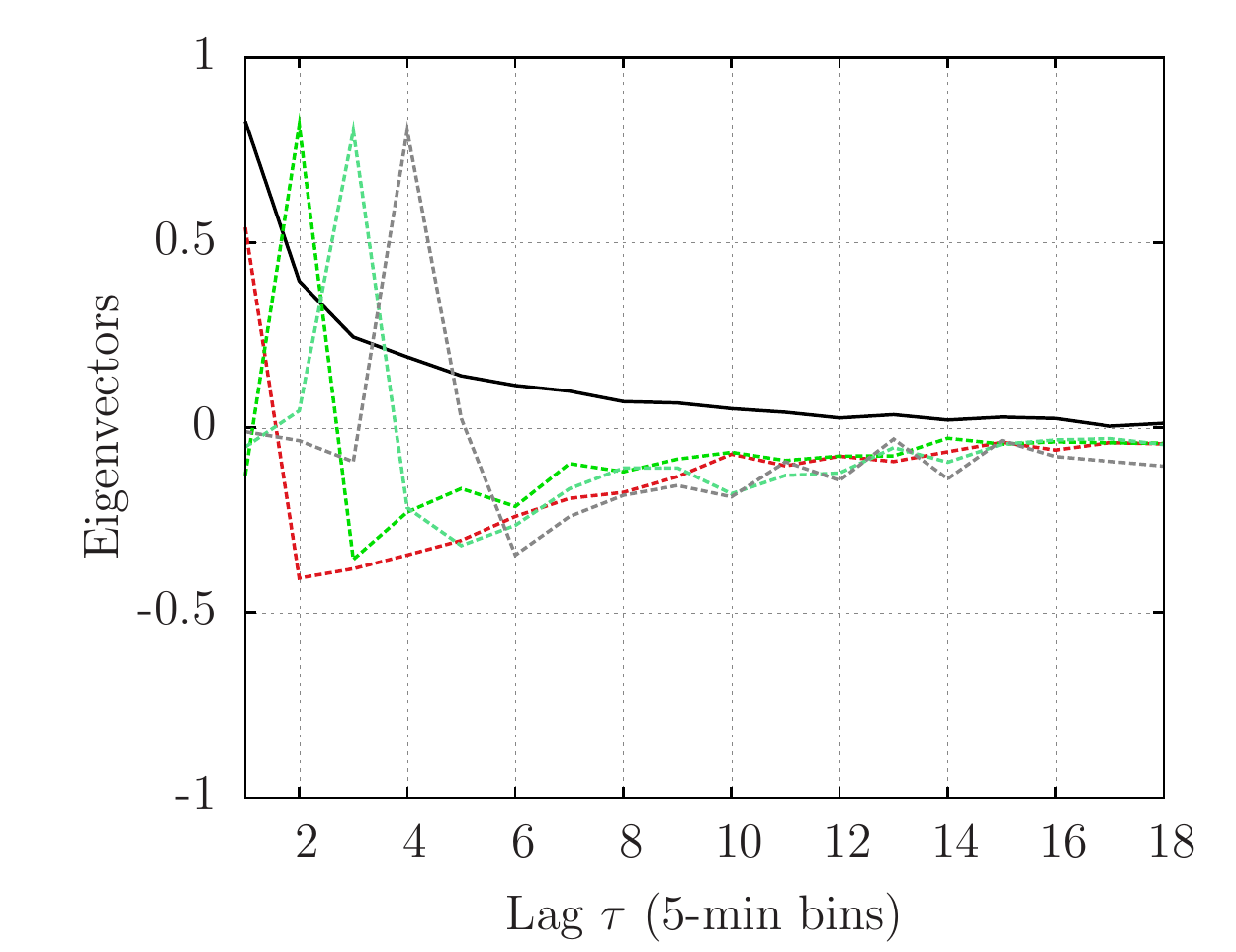}
		\caption{Spectral decomposition of the quadratic QARCH kernel. Left: ranked eigenvalues (plain dark line) and diagonal coefficients (dashed).
		One can see that the diagonal coefficients are very close to the eigenvalues, except for the first eigenvalue which is significantly larger
		than the maximum of the diagonal.
		Right: eigenvectors corresponding to the five largest eigenvalues. The first eigenvector (plain dark line) is a positive decaying kernel, the others
		are close to the canonical vectors $e_i(\tau) = \delta_{i-\tau}$.}
        \label{graph:eigen_QARCH_US_stocks} 
\end{figure}
The spectral decomposition of quadratic kernel (see Figure~\ref{graph:eigen_QARCH_US_stocks}) suggests that $K$ is the superposition of a positive
rank-one matrix and a diagonal one. Indeed, we obtain to a good approximation (see Figure~\ref{graph:ZHfit_QARCH_stocks})
\begin{equation}
K(\tau,\tau') \ \approx \ \phi(\tau) \delta_{\tau-\tau'} + k(\tau) k(\tau')
\nonumber
\end{equation}
where
\begin{equation}
\phi(\tau) = g \tau^{-\alpha}
\quad , \quad
k(\tau) = k_0 \exp(-\omega \tau),
\nonumber
\end{equation}
with $g = 0.09, \ \alpha = 0.60, \ k_0 = 0.14, \ \omega = 0.15$. Note that $\omega=0.15$ corresponds to a characteristic time of about thirty minutes (bin size $ \times \omega^{-1}$) for the decay of the off-diagonal component.
We then fix the off-diagonal part of the kernel $K$ to its fitted value $k(\tau) k(\tau') = k_0^2 \exp(-\omega(\tau+\tau'))$, and we recalibrate the
diagonal of $K$ with a longer maximum lag of $60$ bins (five hours of trading). We obtain
\begin{equation}
\phi_\text{lr}(\tau) = g' \tau^{-\alpha'}
\nonumber
\end{equation}
with the new coefficients $g' = 0.09, \ \alpha' = 0.76$, not far from those obtained above on a shorter time span. 
\begin{figure}[!t]
 \center$
		\begin{array}{cr}
			\hspace{1.5cm}
		\includegraphics[width = 5.5cm, height = 5.5cm]{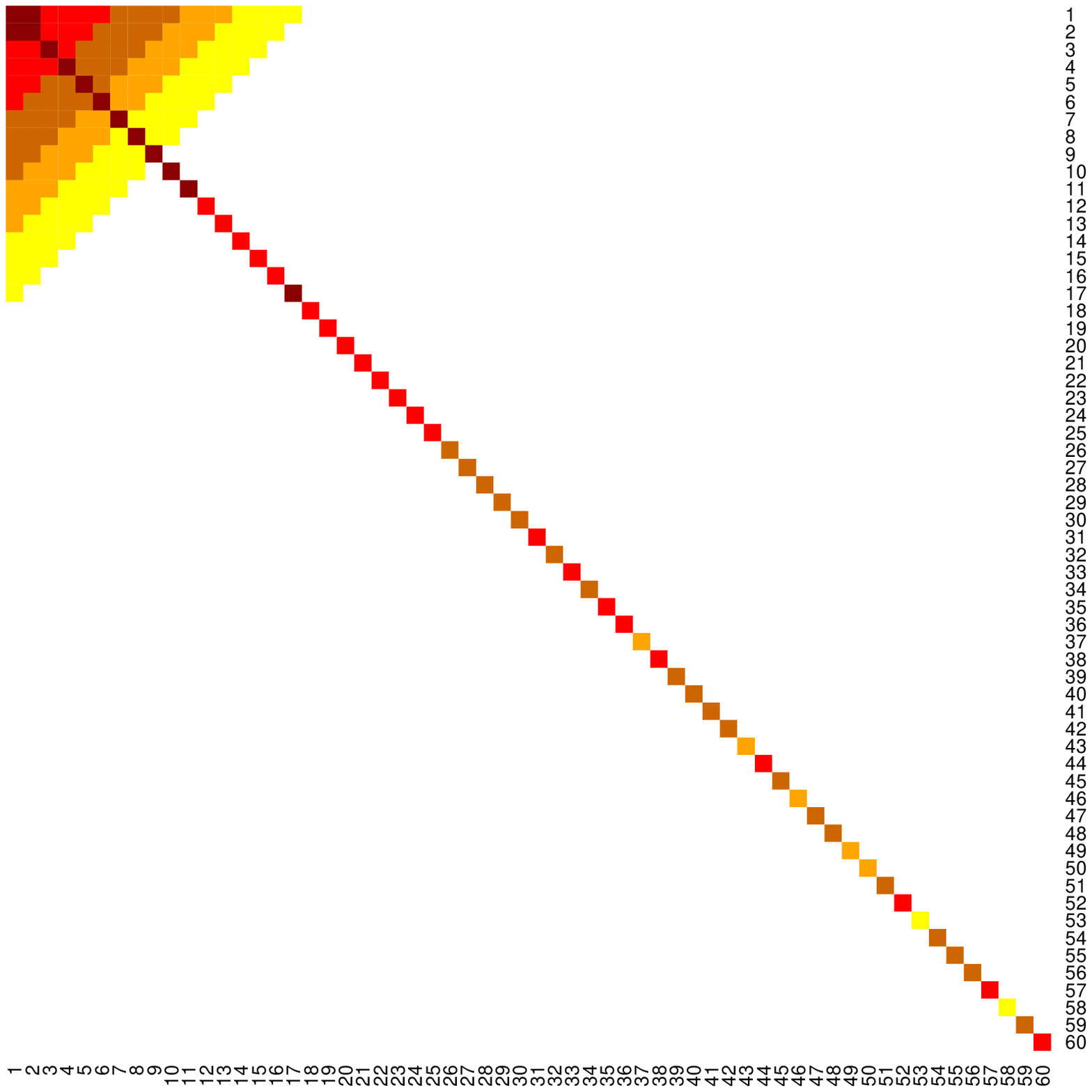}&
		\includegraphics[width = 0.49\textwidth]{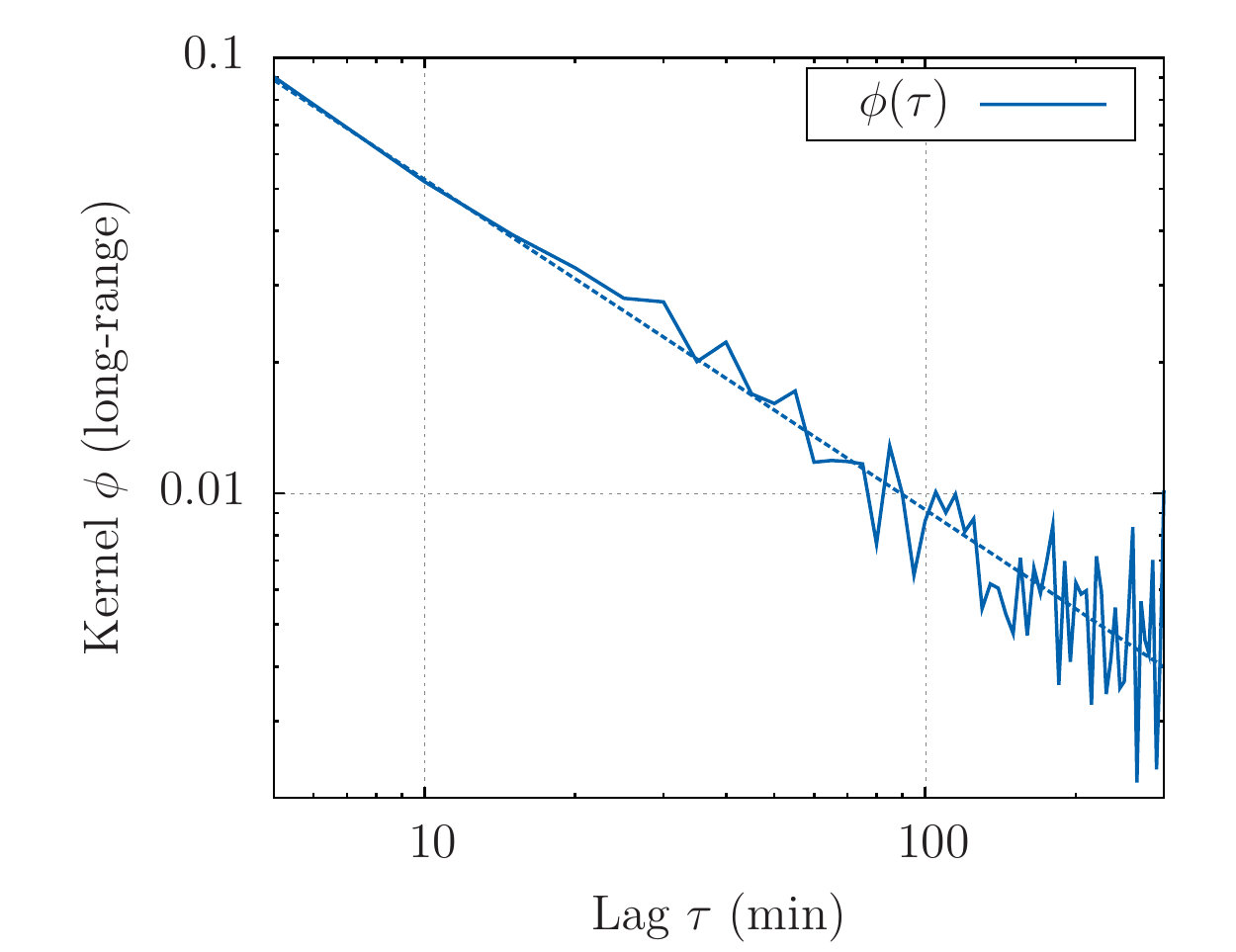}
		\end{array}$
		\caption{Long-range kernel $K$. Left: heatmap of the long-range kernel, with the off-diagonal fixed as its exponential rank-one fit, and with the 
		diagonal calibrated with no constraints. Right: Hawkes kernel $\phi(\tau) = K(\tau,\tau)-k^2(\tau)$ fitted on 60 bins. The kernel $\phi(\tau)$ 
		is plotted in log-log scale, with its power-law fit with exponent $\alpha'=0.76$ (dashed).}
        \label{graph:LR_QARCH_stocks} 
\end{figure}
The residuals $\xi_t$ of the QARCH model, defined by
\begin{equation}
r_t = \sigma_t \xi_t,
\nonumber
\end{equation}
where $r_t$ is the five-minute return and $\sigma_t$ is the QARCH volatility, are modeled with Student's t-distribution.
The calibration of the model with $K(\tau,\tau') = \phi(\tau) \delta_{\tau-\tau'} + k(\tau) k(\tau')$
yields $\nu\approx 7.9$ degrees of freedom for the residuals, which gives a kurtosis $\kappa \approx 4.5$. This has to
be compared with the tail exponent $\nu_r$ of $r_t$ itself, which is, as is well known, in the range $3 \to 4$, see also Fig. \ref{graph:Tails_Data_Hawkes_ZH} below.
Since $\nu$ is more than twice $\nu_r$, the QARCH model with Gaussian residuals and this specific form of $K$ accounts, to a good extent, for the fat tails of five-minute returns, 
that appear to be nearly entirely induced by the quadratic feedback mechanism. We will
justify theoretically and numerically why this is the case in Sections \ref{ZHawkes-exp} and \ref{TRA_ZHawkes}.

In the QARCH model, the endogeneity ratio of the volatility (i.e. the proportion of the volatility that stems from feedback effects) is given by the trace $\text{Tr}(K)$ of the quadratic kernel.
With our parametrization and a maximum lag of $q \geq 1$,
one has
\begin{equation}
\text{Tr}(K) = \overset{q}{\underset{\tau=1}{\sum}} \phi(\tau) + \overset{q}{\underset{\tau=1}{\sum}} k^2(\tau).
\nonumber
\end{equation}
We use the fits $k(\tau) = k_0 \exp(-\omega \tau)$ and $\phi_\text{lr}(\tau) = g' \tau^{-\alpha'}$ to compute $\text{Tr}(K)$ for
$q=78$, which is the total number of five-minute bins in a trading day. We obtain
\begin{equation}
\overset{q}{\underset{\tau=1}{\sum}} \phi(\tau) \simeq 0.74,
\quad
\overset{q}{\underset{\tau=1}{\sum}} k^2(\tau) \simeq 0.06
\quad \Rightarrow \quad
\text{Tr}(K) \simeq 0.80.
\nonumber
\end{equation}
This endogeneity ratio implies that $80\%$ of the intra-day volatility is due to endogenous feedback effects. 
Interestingly, it is close to the value obtained for QARCH and ARCH models at a daily time scale, see \cite{chicheportiche2014fine} and \cite{blanc2014fine}. 
Note that although high, the endogeneity ratio is significantly below the critical limit $\text{Tr}(K)=1$, which is the value found by calibrating a standard 
linear Hawkes process to activity data on much longer time horizons: Ref.~\cite{hardiman2013critical,hardiman2014branching} report $n_H \approx 0.9$ on a time window of a few
hours, and $n_H \approx 0.99$ when the kernel is extended to 40 days. We discuss this issue further in Section~\ref{section:LF_asympt}.

\section{Volatility distribution in the ZHawkes model}
\label{ZHawkes-exp}

\subsection{SDE in the exponential case}

If the kernels $\phi(.)$ and $k(.)$ of the ZHawkes model have an exponential form, the process is Markovian and one can write a 
stochastic differential equation to describe its evolution. Although this assumption is only justified for $k$, this case allows one 
to gain a good intuition on the model, so we investigate this limit in details. It also turns out that the Markovian case is actually extremely 
interesting mathematically.

For the sake of simplicity, let us assume that the price jumps are binary $\xi = \pm \psi$, and we set $\psi=1$ without loss of generality. 
Besides, we note $k(t) = \sqrt{2 n_Z \omega} \, \exp(- \omega t)$ and $\phi(t) = n_H \beta \exp(-\beta t)$, where $n_H$ is the 
Hawkes norm and $n_Z$ the Zumbach norm. We require:
\begin{equation}
\text{Tr}(K) = n_H + n_Z < 1.
\nonumber
\end{equation}
Then the model can be written in this case:
$\lambda_t = \lambda_\infty + H_t + Z_t^2$ where
\begin{equation}
\left \{
\begin{array}{ll}
\dd H_t &= \beta \ \left[ - H_t \ \dd t \ +  n_H  \ \dd N_t \right],
 \\
\dd Z_t &= - \omega \ Z_t \ \dd t \ + \ k_0\ \dd P_t\\
\end{array}
\right.
\label{eqn:Markovian_SDE}
\end{equation}
The processes $N$ and $P$ jump simultaneously with intensity $\lambda_t$ and amplitudes $\Delta N_\tau = 1$ and $\Delta P_\tau = \pm 1$ with equal probability. Although quite simple, this system of jump SDEs lacks tractability compared to a continuous diffusion. 
Thus, we turn to the low-frequency asymptotics that one obtains as the number of jumps in a given time window becomes large, while their amplitudes are scaled down accordingly. This is the object of the following section.

\subsection{Low-frequency asymptotics}\label{section:LF_asympt}

The low-frequency asymptotics of nearly critical Hawkes processes with short-ranged kernels have been investigated in details by Jaisson and Rosenbaum \cite{jaisson2013limit,jaisson2015rough}. They show that for suitable scaling and convergence to the critical point $n_H=1$, the short memory Hawkes-based price process of Bacry et al.~\cite{bacry2014hawkes} converges towards a Heston process (since the Hawkes intensity converges towards a CIR volatility process). The same authors \cite{jaisson2015rough} show that when the kernel exhibits power-law behaviour $\phi(t)\sim t^{-1-\epsilon}$ with $1/2 < \epsilon < 1$, the limiting process for the intensity is a fractional Brownian motion with Hurst exponent $H=\epsilon-\frac{1}{2}$. When $\epsilon$ is close to $1/2$, as empirical data suggests \cite{hardiman2013critical}, the roughness of the latter process is in agreement with the empirical results of \cite{bacry2001modelling,gatheral2014volatility} who find a Hurst exponent $H$ close to zero the \emph{log-}volatility ($H=0$ for the multifractal model of \cite{bacry2001modelling}). However, it is unclear how the Hawkes process intensity can be identified with the log-volatility. A fat-tailed behaviour cannot be reproduced by a simple, linear Hawkes process, as it is absent from Heston-CIR processes (see also below).

Here, we want to investigate the low-frequency asymptotics of the Markovian ZHawkes model, which, as we shall see, reveals very interesting new features, induced by quadratic feedback effects.

Choosing a time scale $T>0$ that will eventually diverge, we define the processes $\bar{H}_t^T = H_{tT}$, $\bar{Z}_t^T = Z_{tT}$, $\bar{N}_t^T = N_{tT}$ and $\bar{P}_t^T = P_{tT}$, with the parameters $\beta_T$ and $\omega_T$ that may depend on $T$, but with fixed endogeneity parameters $n_H$ and $n_Z$:
Equation~\eqref{eqn:Markovian_SDE} gives
\begin{equation}
\left \{
\begin{array}{ll}
\dd \bar{H}^T_t &= -\beta_T \left[\ \bar{H}^T_t \ T \dd t \ + \ n_H \ \dd \bar{N}^T_t \right],\\
\dd \bar{Z}^T_t &= - \omega_T \ \bar{Z}^T_t \ T \dd t \ + \ \gamma_T \ \dd \bar{P}^T_t,  \\ 
\end{array}
\right.
\end{equation}
where $ \gamma_T^2 := 2 \omega_T n_Z$ and the common jump intensity of $\bar{N}^T$ and $\bar{P}^T$ is $T \times [\lambda_\infty + \bar{H}^T_t + (\bar{Z}^T_t)^2 ]$. 
Since the signs of the jumps of $\bar{P}^T$ are assumed to be unpredictable and equal to $\pm 1$, the infinitesimal generator of the process is given by
\begin{align}
\mathcal{A}^T f(h,z) \ &= \ - \beta_T \ h \ T \ \partial_h f(h,z) 
\ - \ \omega_T \ z \ T \ \partial_z f(h,z) 
\label{eqn:generator_T} \\
& \qquad + T \left[\lambda_\infty+h+z^2\right] 
\left\{ \frac12 f\left ( h+ n_H \beta_T, \ z+\gamma_T \right ) +  \frac12 f\left ( h+ n_H \beta_T, \ z-\gamma_T \right ) - f\left (h,z \right )  \right\}
\nonumber
\end{align}
for any functions $f$ twice continuously differentiable on $(0,+\infty) \times \mathbb{R}$.
We now consider the following scaling
\begin{equation}
\ \beta_T = \overline{\beta}/T, \qquad, \ \omega_T = \overline{\omega}/T,
\label{eqn:spacial_scaling}
\end{equation}
with $\overline{\beta},\overline{\omega}>0$. Since we fixed the values of $n_H$ and $n_Z$, our procedure can be called a \enquote{constant endogeneity rescaling},
as opposed to the scaling used by Jaisson and Rosenbaum in~\cite{jaisson2013limit} and~\cite{jaisson2015rough}, where the endogeneity ratio $n_H$ of the process needs to converge to unity as $T$ goes to infinity. Our choice is partly motivated by the calibration results of Section~\ref{section:QARCH_calib} for intra-day returns, that yield an endogeneity ratio in the range $0.7-0.9$, close to what is obtained at the daily time scale in~\cite{chicheportiche2014fine} and~\cite{blanc2014fine}, and significantly away from the critical value $n_H=1$. Equations~\eqref{eqn:generator_T} and~\eqref{eqn:spacial_scaling} then combine as
\begin{align}
\mathcal{A}^T f(h,z) \ &= \ - \overline\beta \ h \ \partial_h f(h,z) 
\ - \ \overline\omega \ z \ \partial_z f(h,z) 
\nonumber \\
& \ + T\left[\lambda_\infty+h+z^2\right]  T 
\left\{ \frac12 f\left ( h+n_H \frac{\overline\beta}{T}, \ z+\frac{\overline\gamma}{\sqrt{T}} \right ) + 
 \frac12 f\left ( h+n_H \frac{\overline\beta}{T}, \ z-\frac{\overline\gamma}{\sqrt{T}} \right ) - f\left (h,z \right )  \right\},
\nonumber
\end{align}
where we introduced $\overline\gamma=\sqrt{2n_Z \overline\omega}$. We turn to the low-frequency asymptotics. As $T$ goes to infinity, one has
\begin{equation}
\frac12 f\left ( h+ n_H\frac{\overline\beta}{T}, \ z+\frac{\overline\gamma}{\sqrt{T}} \right ) + 
 \frac12 f\left ( h+ n_H \frac{\overline\beta}{T}, \ z-\frac{\overline\gamma}{\sqrt{T}} \right ) - f\left (h,z \right ) 
=
\frac{n_H \overline\beta}{T} \partial_h f(h,z) + \frac{\overline{\gamma}^2}{2T} \partial^2_{zz} f(h,z)
+ \ \text{o}\left( \frac1 T\right),
\nonumber
\end{equation}
therefore $\mathcal{A}^T f(h,z)$ converges to
\begin{equation}
\mathcal{A}^\infty f(h,z)
\ = \
- \overline\beta \left[ (1-n_H) h - n_H (\lambda_\infty+z^2) \right] \partial_h f(h,z) 
\ - \ \overline\omega z \partial_z f(h,z) 
\ + \ n_Z \overline\omega \left[\lambda_\infty+h+z^2\right] \partial^2_{zz} f(h,z).
\nonumber
\end{equation}
The operator $\mathcal{A}^\infty$ is the infinitesimal generator of the diffusion
\begin{equation} \label{limit-process}
\left \{
\begin{array}{ll}
\dd \bar{H}^\infty_t &= \left[ - (1-n_H) \ \bar{H}^\infty_t 
+ n_H \left(\lambda_\infty+\left( \bar{Z}^\infty_t \right)^2\right) \right] \overline\beta \dd t, \\
\dd \bar{Z}^\infty_t &= - \overline\omega \ \bar{Z}^\infty_t \ \dd t 
\ + \ \overline\gamma \ \sqrt{\lambda_\infty+\bar{H}^\infty_t+\left( \bar{Z}^\infty_t \right)^2} \dd W_t,\\
\end{array}
\right.
\end{equation}
where $W$ is a standard Brownian motion. A standard argument of Kallenberg~\cite{kallenberg} (Theorem 19.25) then gives the convergence of the process $(\bar{H}^T,\bar{Z}^T)$ to  $(\bar{H}^\infty,\bar{Z}^\infty)$ as $T$ goes to infinity. Hence, one does \emph{not} need that the norm of the process tends to 1 (i.e. that the process is nearly critical) for a non-degenerate limit process to be obtained. The above limiting process is the major result of this section. Although it was derived
for a Markovian ZHawkes process, we believe that this is the limiting process for the whole class of non-critical ZHawkes processes with short memory, and is the analogue of the Heston-CIR limiting process for Hawkes, as in \cite{jaisson2013limit}. The limiting behaviour corresponding to long-memory/critical ZHawkes processes, in the spirit of \cite{jaisson2015rough}, is left for future investigations. We now investigate some of the properties of the limiting process, Eq. (\ref{limit-process}), in particular the induced tail of the volatility distribution.

\subsection{Tail of the volatility distribution}\label{section:tails_LF}

From now on we drop the superscript $\infty$ on $\bar{H}$ and $\bar{Z}$; the fact that we are studying the limiting process is implied.
Let us note first that there is no Brownian part in the SDE for $\bar{H}$ so that it can be solved explicitly as a deterministic function of $(\bar{Z}_s)_{s\leq t} \ $ :
\begin{equation}
\bar{H}_t \ =  \ \bar{H}_\infty
\ + \  n_H \overline\beta \int_{-\infty}^t \exp(-(1-n_H)\overline\beta  (t-s)) 
 \bar{Z}_s^2 \dd s; \qquad \bar{H}_\infty :=\frac{\lambda_\infty}{1 - n_H}
\nonumber
\end{equation}
In the considered limit, $\bar{H}_t$ can thus be written as the sum of a constant term and an exponential moving average of the square of $\bar{Z}_s$. 
We get the autonomous, but non-Markovian SDE for $\bar{Z}_t$:
\begin{equation}
\dd \bar{Z}_t = - \overline\omega \ \bar{Z}_t \ \dd t 
\ + \ \overline\gamma \ \sqrt{ \bar{H}_\infty
+ \bar{Z}_t^2
+ n_H \overline\beta \left[\int_{-\infty}^t \exp(-(1-n_H)\overline\beta (t-s)) 
\bar{Z}_s^2 \dd s \right]
} \ \dd W_t.
\label{eqn:SDE_whawkes}
\end{equation}

\subsubsection{ZHawkes without Hawkes}

We first consider the simpler case where the Hawkes feedback is zero, i.e. $n_H=0$. This corresponds to the case where only the Zumbach term is present
in the starting model, i.e. $\lambda_t = \lambda_\infty + Z_t^2$ in Equation~\eqref{eqn:ZHawkes_model}. As we see in the sequel, this simpler model is still rich enough to reproduce some interesting empirical properties of the volatility process. One gets:
\begin{equation}
\dd \bar{Z}_t = - \overline\omega \ \bar{Z}_t \ \dd t 
\ + \ \overline\gamma \ \sqrt{\lambda_\infty + \bar{Z}_t^2} \ \dd W_t,
\label{eqn:SDE_nohawkes}
\end{equation}
which is a particular case of Pearson diffusions, which are extensively described and classified by Forman and Sorensen~\cite{forman2008pearson}.
The process $\bar{Z}/\sqrt{\lambda_\infty}$ fits in Case 3 of their classification (see~\cite{forman2008pearson} Section 2.1), with the dictionary
$\mu \to 0, \theta \to \overline\omega$ and $a \to n_Z$. Therefore, $\bar{Z}_t$ is ergodic and its stationary law is a Student t-distribution with $1+1/n_Z$ degrees of freedom and scale parameter $\sqrt{n_Z\lambda_\infty/(1+n_Z)}$.  This implies that stationary law of the square of $\bar{Z}^\infty$ is a F-distribution with $1$ and $1+1/n_Z$ degrees of freedom, and scale parameter $n_Z\lambda_\infty/(1+n_Z)$.
We will denote as
\begin{equation}
V_t \ = \ \psi^2 \left[\lambda_\infty + \bar{Z}_t^2 \right]
\nonumber
\end{equation}
the low-frequency squared volatility of the price (we reintroduced the jump size $\psi$ for completeness). A straightforward change of variables yields the stationary density $q(v)$ of the process $V$ as:
\begin{equation}
q(v) = \frac{\Gamma\left( 1+\frac1{2n_Z} \right)}{\Gamma\left( \frac12+\frac1{2n_Z} \right) \sqrt{\pi v_{\infty}(v-v_\infty)}}
\left( \frac{v_{\infty}}{v} \right)^{\left(1+\frac1{2n_Z}\right)} \mathbbm{1}_{\{v>v_{\infty}\}}
\label{eqn:distrib_V_nohawkes}
\end{equation}
where $v_{\infty} = \lambda_\infty \psi^2$ is the baseline level of the squared volatility. 
For the tail exponent of the distribution of $V_t$, we get a power-law tail:
\begin{equation}
q(v) \ \underset{v \rightarrow +\infty}{\sim} \
C \ v^{-\left(\frac32+\frac1{2n_Z}\right)}
\label{eqn:tail_V_nohawkes}
\end{equation}
with $C$ an explicit constant.
We find this result interesting for two reasons. First, one obtains a power-law behavior that emerges naturally from the fact that since the volatility behaves as $|\bar{Z}_t|$ for large values of $\bar{Z}_t$, the process describing its dynamics is simply a multiplicative Brownian motion with drift (see \ref{eqn:SDE_nohawkes}). This is at variance with the \enquote{diagonal} Hawkes counterpart of \cite{jaisson2013limit} where the coefficient in front of the 
Brownian noise is only the {\it square-root} of the volatility, which inevitably leads to a process that has a characteristic scale and thin tails. 
Second, the stationary distribution of $V$ only depends on the Zumbach norm $n_Z$, that can be seen as the endogeneity of the process. This last result suggests that, similar to Hawkes processes 
where the asymptotic properties only depend on the norm $n_H$ as soon as the kernel is short-ranged, the distribution~\eqref{eqn:distrib_V_nohawkes} of the squared volatility should hold for any short-ranged kernel.

Another remark is that as soon as $n_Z\geq1/3$, the variance $\sigma_V^2$ of the activity $V$ explodes while its mean $\mu_V$ remains finite up to $n_Z \to 1^-$. Now, when fitting the time 
series generated by this process using a simple Hawkes process, one finds $n_H \approx 1 - \sqrt{\mu_V(W)/\sigma_V^2(W)}$ for a suitable choice of window size $W$ (see~\cite{hardiman2014branching}). Therefore, the vanishing of the mean/variance ratio necessarily imposes that the fitted Hawkes process must be critical, i.e. $n_H=1$! What we argue here is that this {\it apparent} criticality may in fact be induced by quadratic feedback effects, but does not necessarily imply that the true underlying process is critical.

Finally, note that in the diffusive limit where the price process satisfies the equation $\dd \bar{P}^\infty_t = \sqrt{V_t}\dd W_t$, the asymptotic stationary distribution for the returns is given by:
\begin{equation}
p(r) \tail{|r|}{\infty} \frac{C'}{|r|^{1+\nu}}; \qquad \nu \equiv 1+\frac{1}{n_Z}.
\nonumber
\end{equation}
The fat-tail volatility that is generated by our model naturally produces a fat-tail distribution of instantaneous returns, with exponent $\nu$ for the 
cumulative distribution equal to $1+1/n_Z \geq 2$. 
The more endogenous, the fatter the tails for the returns: this interpretation seems intuitive. For a critical process, $n_Z=1$, the tail is such that
the volatility of the returns diverges. A tail exponent for the cumulative distribution $\nu \approx 3$ (the so-called ``inverse cubic law'', observed on a large universe of traded products) is obtained for $n_Z=0.5$. Note however that the value of $n_Z$ obtained above from calibrating the model is much smaller, $n_Z \approx 0.06$, leading to $\nu \approx 18$, far too large to explain the tail of financial returns. We will see now that, quite interestingly, the interaction with a non-critical Hawkes kernel can substantially reduce the value of $\nu$.

\subsubsection{ZHawkes with Hawkes}

The case when $n_H > 0$ is more complicated but, remarkably, the tail exponent of the activity distribution $q(v)$ can still be analytically computed in some limits. 
The idea is to realize that when $\bar{Z} \to \infty$, the distribution of $\bar{H}$ conditional to a certain large value of $\bar{Y}:=\bar{Z}^2$ is of the form:
\begin{equation}
\Pi({\bar{H}}|{\bar{Y}}) = \frac{1}{\bar{Y}} F\left(\frac{H}{\bar{Y}}\right) + o(\bar{Y}); \qquad (\bar{Y} \to \infty),
\nonumber
\end{equation}
where $F(.)$ is a certain scaling function which obeys a differential equation derived in Appendix B. Correspondingly, one can 
show that the far-tail of the distribution of $V_t \ = \ \psi^2 \left[\lambda_\infty + \bar{Z}_t^2 \right]$ is still a power-law, given by:
\begin{equation}
q(v) \ \underset{v \rightarrow +\infty}{\sim} \ C'' \ v^{-\left(\frac32+\frac{1}{2n_Z(1+a^*)}\right)},
\label{eqn:tail_V_hawkes}
\end{equation}
where $C''$ is another constant and $a^*$ is defined as:
\be
a^* = \int_0^\infty {\rm d}x \, \, x \, F(x).
\ee
Introducing $\chi:= \frac{2 \overline\omega}{\overline \beta}$ as the ratio of the correlation time scale of the Hawkes process to the one of the ZHawkes process, a full solution for 
$F$ can be found in the two limits $\chi \to 0$ and $\chi \to \infty$, allowing one to fix the value of $a^*$. One finds (see Appendix B):
\be
a^* \approx \frac{n_H}{1-n_H} \left[ 1 - \chi \frac{1-n_H-n_Z}{(1-n_H)^2} \right], \quad (\chi \to 0); \qquad a^* \approx \frac{n_H}{\chi (1-n_Z)}, \quad (\chi, \chi n_Z \to \infty).
\ee
Two other limiting cases can be exactly solved: one is when $n_H \to 0$, one finds that $a^* \approx \frac{n_H}{\chi (1-n_Z)}$ still holds provided $a^* \ll 1$, 
and the other is $n_Z \to 0$, for which we find an explicit expression for $a^*$ as the solution of a second degree equation (see Appendix B).   

The corresponding exponent for the asymptotic tail of the cumulative distribution of returns is now given by: 
\begin{equation}
\nu = 1 + \frac{1}{n_Z(1+a^*)}, 
\end{equation}
with:
\begin{itemize}

\item for $n_H = 0$ (ZHawkes without Hawkes), one recovers the previous case where $a^* = 0$ and $\nu = 1 + 1/n_Z$.

\item for $0 < n_H <  1$ and $\chi \to 0$ (Hawkes much ``faster'' than ZHawkes), the exponent $\nu$ is {\it decreased} to $\nu = 1 + (1 - n_H)/n_Z + O(\chi)$.

\item for $0 < n_H <  1$ and $\chi \to \infty$ (Hawkes much ``slower'' than ZHawkes), the exponent $\nu$ is again {\it decreased} from $\nu = 1 + 1/n_Z$ 
by an amount $\sim 1/\chi$. 

\item In the case $n_Z \to 0$, one finds $\nu = 1 + \frac{b}{n_Z}$, where $b$ can be computed in terms of $n_H$ and $\chi$, see Appendix B.
 
\end{itemize}

The results of this section are, we believe, quite interesting. First, the two-dimensional limit process defined by Eqs. (\ref{limit-process}) leads to power-law tails for the volatility 
that can be exactly characterized in some limits. From a theoretical point of view, the possibility of computing exactly the tail exponent in this model is potentially 
important if our ZHawkes process turned out to be a central ingredient to model the dynamics of financial markets. Second, we have found that although the Hawkes kernel per-se does not lead to power-law tails (i.e., $\nu \to \infty$ when $n_Z \to 0$), the Hawkes kernel actually ``cooperates'' with the ZHawkes kernel to make the tails of the distribution fatter. The case of empirical interest is $n_Z = 0.06$, $n_H \approx 0.8$ leads to $\nu = 1+ (1-n_H)/n_Z \approx 4$ for $\chi \to 0$, which indeed remains in the experimental range for a non-Markovian ZHawkes process with parameters calibrated on intraday data, as will be shown by numerical simulations in the next section. 

We find this phenomenon quite remarkable: whereas the Hawkes feedback alone is not able to explain fat-tails, only a relatively small amount of quadratic (Zumbach) feedback 
generates power-law tails in the correct range (remember that $n_Z = 0.06 \ll n_H$).
Note however that this ZHawkes family of models leads a continuously varying exponent (as a function of the parameters) rather than a 
fixed, universal exponent like in many physical situations. This begs the question: is there any mechanism that would explain why the feedback parameters 
$n_Z, n_H, \chi$ lie in a rather restricted interval, such as to explain the apparent universality of the tail exponent of (mature) financial markets?

\section{Numerical simulation results}\label{TRA_ZHawkes}

\subsection{Empirical tails of the volatility process}

In this section, we compare numerically the volatility process generated by the ZHawkes model, with a standard Hawkes-based price model and with the financial data studied in Section~\ref{section:dataset}. We simulate a ZHawkes model with an exponential Zumbach part and a power-law Hawkes part, with parameters inspired by the QARCH calibration of Section~\ref{section:QARCH_calib}: for $t$ expressed in minutes,
\begin{equation}
\phi(t) = 0.0016 \times (1+0.01 \times t)^{-1.2},
\quad
k(t) = 0.003 \times \exp(-0.03 \times t),
\nonumber
\end{equation}
so that $n_H = 0.8$, $n_Z = 0.1$ and $\text{Tr}(K) = 0.9$. Note that to simulate a stationary ZHawkes model, we choose a decay exponent above $1$ for $\phi$, although the QARCH calibration suggests 
a slower decay for $t$ corresponding to intraday time scales. Although not fully satisfactory, this is the simplest way to enforce stationarity without having to introduce a more complicated 
functional form for $\phi(t)$ that would model overnight effects and daily time scales. As a benchmark, we also simulate a standard Hawkes-based price process ($n_Z \equiv 0$) with $\phi = (1+0.01 \times t)^{-1.3}$, $n_H=0.99$, which is close to the calibration results of~\cite{hardiman2013critical}.

It is important to note that to simulate the ZHawkes and the Hawkes model, we choose constant price jumps $\Delta P_\tau = \pm \psi$. Therefore, our numerical results for the distribution of the volatility can by no means be attributed to the kurtosis of individual price jumps.

\begin{figure}[!ht]
\center
\includegraphics[height=7cm]{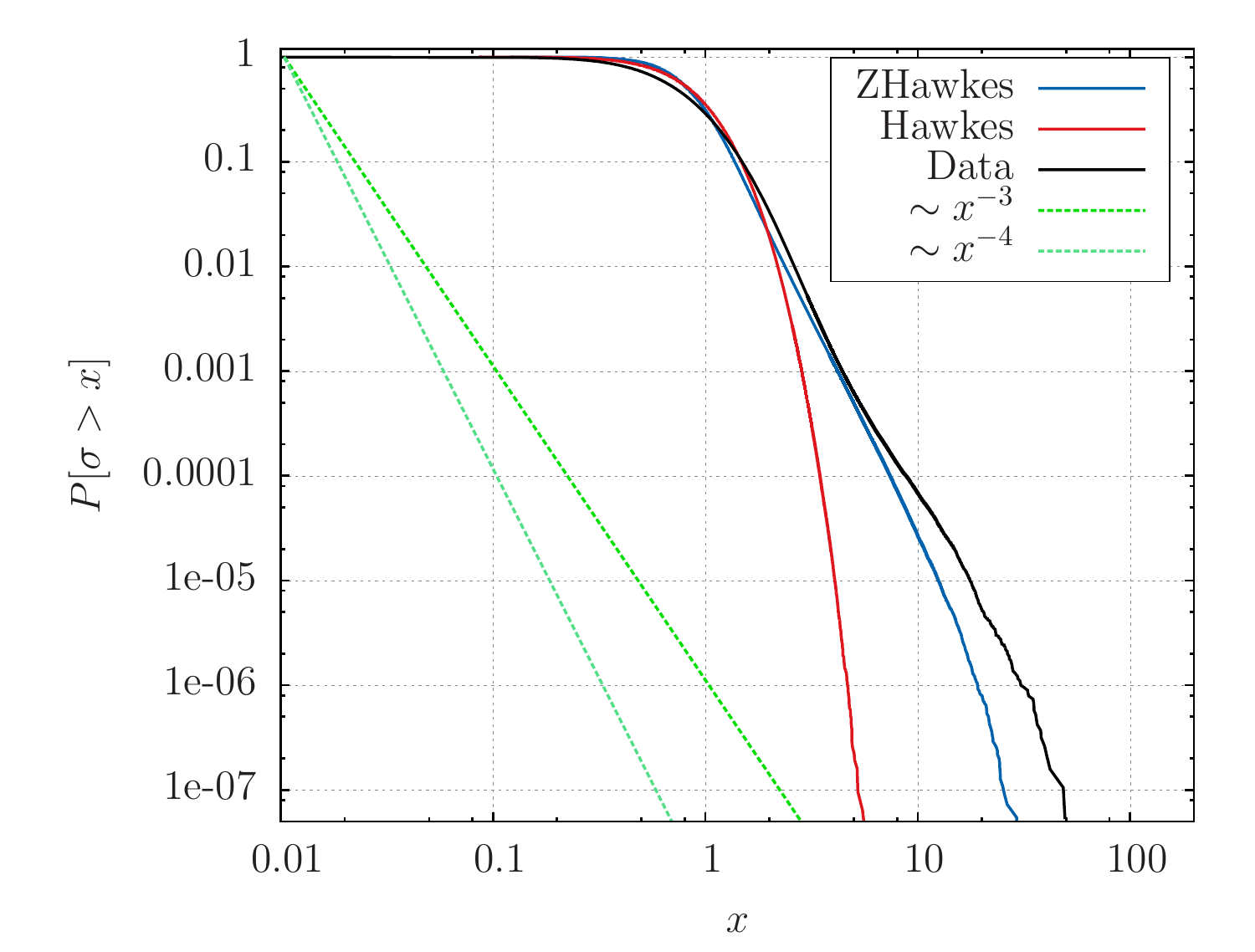}
\caption{Cumulative density function of the Rogers-Satchell volatility for US stock data (plain line), simulated Hawkes data (red dashed line),
and simulated ZHawkes data (blue dot-dashed line). Notice how well the empirical distribution function is reproduced by the ZHawkes model, calibrated as in 
Section \ref{sec:calibration}.}
\label{graph:Tails_Data_Hawkes_ZH} 
\end{figure}

\begin{figure}[!ht]
\center
\includegraphics[width=\textwidth]{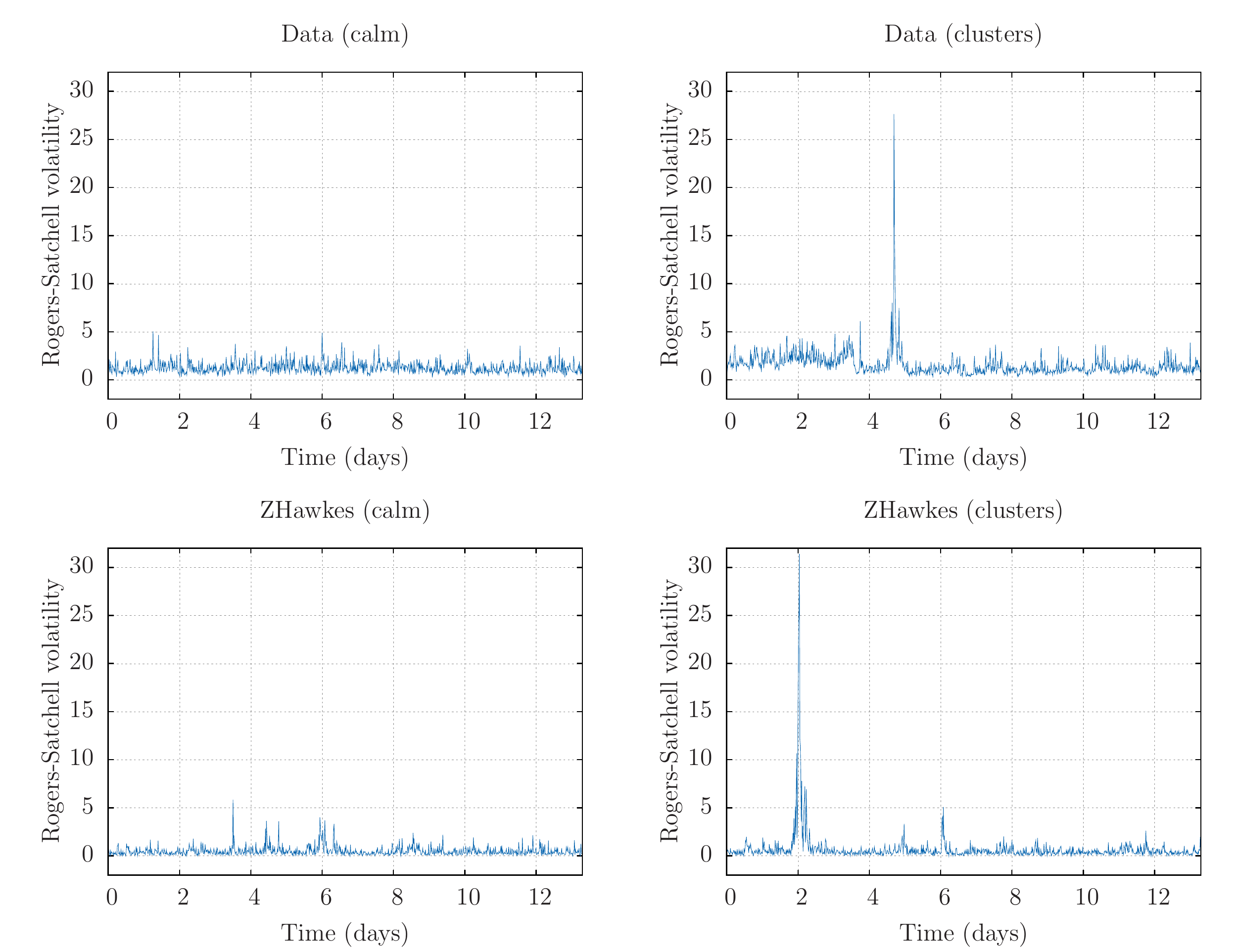}
\caption{Time series of Rogers-Satchell volatility. Above: real data ; below: simulated ZHawkes data ; left: period of calm ;
right: cluster of intense activity.}
\label{graph:Vol_paths_Data_ZH} 
\end{figure}

For both simulated and real data, we consider the Rogers-Satchell volatility times series for five-minute bins.
We use the Hill exponent~\cite{hill1975simple} as an estimator of the empirical tail exponent of the volatility
\begin{equation}
\nu_{\text{hill}} = 1+\frac1{\frac1n \sum_{i=1}^n \log(\sigma_i/\sigma_\text{min})}
\nonumber
\end{equation}
where $\sigma_\text{min}$ is some cutoff and $\sigma_i \geq \sigma_\text{min}$ are the volatilities in the far tail region of the distribution.
One obtains $\nu_{\text{hill}}=4.50$ for the (normalized) five minutes returns of US stocks (in agreement with many previous determinations of this exponent), $\nu_{\text{hill}}=5.07$ for 
the ZHawkes model and $\nu_{\text{hill}}=12.4$ for the standard Hawkes-based model without ZHawkes feedback. Even with a norm close to one and a slowly-decaying kernel, 
the standard Hawkes model cannot reproduce the tails observed on US stock data. Instead, the ZHawkes model, with a norm strictly below unity and a short-lived Zumbach effect, 
naturally produces fat tails very similar to those observed empirically, even with a rather small $n_Z$. These observations are illustrated by Figures~\ref{graph:Tails_Data_Hawkes_ZH} and~\ref{graph:Vol_paths_Data_ZH}.

\subsection{Time-reversal asymmetry of ZHawkes processes}

Another salient feature of financial markets is, as discussed in the introduction, the time-reversal asymmetry (TRA) of the price time series. The authors of~\cite{chicheportiche2014fine} 
study this feature for stock data on the one hand, and for a simulated FIGARCH volatility process 
on the other. The chosen observable is the cross-correlation of present Rogers-Satchell volatilities $\sigma_{t}^2$ with past squared returns $r_{t-\tau}^2$, to that of present squared returns with past volatilities, which is found to be such that $\langle r_{t-\tau}^2 \sigma_{t}^2 \rangle_t > \langle r_{t}^2 \sigma_{t-\tau}^2 \rangle_t$ for $\tau > 0$, both for real data and FIGARCH processes. 

This observation is one of the main motivations for the model introduced in the present paper, since standard models that use Brownian SDEs are TRS by construction and cannot reproduce this asymmetry. In this section, we measure the amount of TRA for the simulated ZHawkes process and for the Hawkes benchmark described in the previous section, and for the financial dataset studied in Section~\ref{section:dataset}.

As in Sections~\ref{section:QARCH_calib}, we consider the returns and the Rogers-Satchell volatilities defined for intra-day five-minute bins. Here, the maximum lag $q$ is fixed to $36$ ($36$ bins of $5$ minutes $= 3$ hours of trading) and the lag index $\tau$ varies between $1$ and $q$. We introduce
\begin{itemize}
\item The cross-correlation function of the Rogers-Satchell volatility and absolute returns
\begin{equation}
C(\tau) = \frac{\langle \volRS_{t} \times|r_{t-\tau}|\rangle - \langle \volRS \rangle \langle |r| \rangle}
{\sqrt{\langle {\volRS}^2 \rangle - \langle \volRS \rangle^2}\sqrt{\langle r^2 \rangle - \langle |r| \rangle^2}}.
\nonumber
\end{equation}
\item The time asymmetry ratio
\begin{equation}
\Delta(\tau) = \frac{\overset{\tau}{\underset{\tau'=1}{\sum}} [C(\tau')-C(-\tau')]}
{2 \overset{q}{\underset{\tau'=1}{\sum}} \max(|C(\tau')|,|C(-\tau')|)}
\in [-1,1].
\nonumber
\end{equation}
\end{itemize}

Note that we choose to compute the cross-correlation function using the \textit{absolute} returns instead of the
\textit{squared} returns, since it yields results that are less noisy and more robust to tail events (and thus less sensitive to
the normalization method).

\begin{figure}[!ht]
\center
\includegraphics[height=7cm]{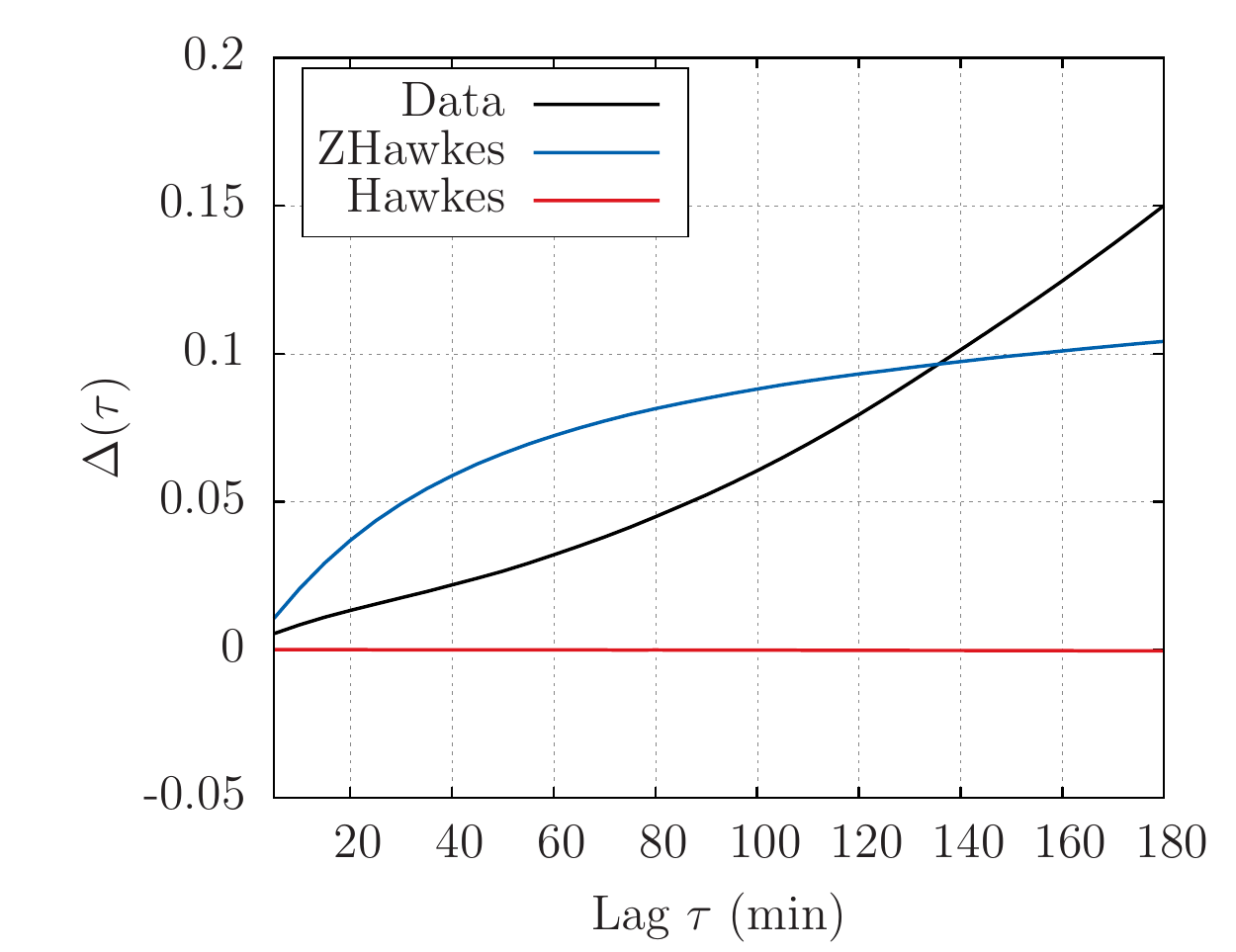}
\caption{Time asymmetry ratio $\Delta(\tau)$ for US stock data (plain line), simulated Hawkes model (red line),
and simulated ZHawkes model (blue dot-dashed line). Note that the Hawkes process does not generate any detectable TRA.}
\label{graph:TRI_Data_Hawkes_ZH} 
\end{figure}

We compare the time asymmetry ratios $\Delta(\tau)$ for real stock returns, returns simulated with the ZHawkes model and returns simulated with a standard Hawkes-based price model.
The results are illustrated by Figure~\ref{graph:TRI_Data_Hawkes_ZH}. The standard Hawkes model, perhaps surprisingly, does not generate any detectable TRA: $|\Delta(\tau)| < 10^{-3}$ for all $\tau$. Thus it is clear 
that the Hawkes model with no off-diagonal quadratic feedback cannot reproduce the time asymmetry observed in intra-day volatility, for which $\Delta(\tau)$ is one hundred times larger.
On the other hand, the ZHawkes model with parameters in line with the QARCH calibration of Section~\ref{sec:intra_QARCH} features some time asymmetry,
which is not only of the correct sign but also reproduces the right order of magnitude, without any further parameter adjustment. However, the function $\tau \mapsto \Delta(\tau)$ is 
found to be concave for the ZHawkes model (as expected on general grounds) and, strangely, convex for stock data. Even with a thorough normalization protocol,
intra-day returns are not rigorously stationary, and we believe that the convexity of $\tau \mapsto \Delta(\tau)$ observed on real data is spurious, as it should should saturate to a value 
less than $1$ beyond some time scale. Such convexity would probably be hard to reproduce with a simple model, unless it some non-stationary is added by hand.

\section{Conclusion}\label{sec:conclusion}

The central message of our study is that the standard Hawkes feedback, where past activity increase the intensity of the current activity, fails at accounting for two essential features of the 
dynamics of markets: a) the fat-tails in the activity/volatility cannot be reproduced and b) the time-reversal asymmetry between past daily volatilities and future intraday volatilities or 
vice-versa is completely absent within the Hawkes framework. This was not a priori obvious, since Hawkes processes are constructed on the idea of a feedback from the past. 
We have thus proposed QHawkes processes as simple, intuitive generalisations of the Hawkes process which posit that the feedback is in fact not only on the past activity, but on past price returns themselves. 

A QHawkes model can be seen as a consistent definition of a Quadratic ARCH (QARCH) model as a continuous-time point-processes. This in fact allowed us to calibrate a QHawkes model on the intraday returns of 133 NYSE stocks. We find that the matrix kernel of the QHawkes has a diagonal part (corresponding to the standard Hawkes component) and a off-diagonal, rank-one part that we call ``ZHawkes''. It
corresponds to Zumbach's insight that local trends in the price, both up or down, generate more future activity. ZHawkes processes have some interesting properties that standard Hawkes processes lack, namely: (i) the quadratic feedback naturally produces a multiplicative dynamics for the volatility, generating power-law tails for the volatility and the returns, (ii) it can generate long memory without necessarily be at its critical point (iii) it reproduces a level of time-reversal asymmetry (TRA) that is fully compatible with what is measured on actual financial data. The continuous limit SDE corresponding to exponential kernels is found to be a tractable two-dimensional generalization of Pearson diffusions. In particular the 
tail exponent of the volatility can be exactly computed in several cases and, quite remarkably, fall within the empirical range even when the ZHawkes kernel is of small amplitude. These mathematically tractable diffusions are reminiscent of the log-normal volatility processes considered in \cite{stein1991stock,bacry2001modelling, bergomi2005smile} and more recently \cite{gatheral2014volatility}, and provide a natural ``microscopic'' mechanism for a multiplicative process for the volatility 
itself, which up to now has remained quite a mysterious hypothesis \cite{jaisson2015rough}. 

We hope our paper motivates more developments on the family of QHawkes models. We have indeed only touched upon the mathematical properties and the empirical relevance of such models but we
believe that deeper work on the subject would be valuable, in particular concerning the precise calibration of the model itself. A completely open question at this stage is the treatment
of overnights and the generalisation of the model to describe longer time scales (our calibration was restricted to intraday data), generalizing the QARCH description proposed by two of us in \cite{blanc2014fine}. In particular, we know that time-reversal asymmetry can still be detected on time scales of days or weeks \cite{zumbach2001heterogeneous,chicheportiche2014fine} and this 
can certainly not be reproduced with a ZHawkes kernel decaying over 30 minutes, as found here. Similarly, multiplicative log-normal models for the volatility have commonly been considered for 
daily returns. How much is the fat-tailed, long memory of the volatility, recently described within the context of standard Hawkes process, should in fact be traced to the QHawkes 
mechanism proposed here is, in our opinion, a very interesting question for future research.

To conclude, we believe that a comprehensive understanding of the volatility process, from the scale of the event up to macroscopic scales, would seem very valuable in several respects, in particular that of market design. One would perhaps understand how a change in market microstructural rules (e.g. the tick size) may affect its macroscopic properties (e.g. volatility). Finding a solid, behavioural microscopic foundations to the volatility process seems crucial: when fully understood, simple constraints on the agents might then change the overall, macroscopic market behaviour. We hope that our generalized Hawkes process could provide some clues on this issue.

\subsection*{Acknowledgments} 

We want to thank R. Chicheportiche, J. Gatheral, S. Hardiman, Th. Jaisson, I. Mastromatteo and M. Rosenbaum for many insightful discussions on these issues.


\nocite{chatterji1998cours}
\nocite{zaanen1956linear}

\bibliographystyle{abbrv}
\bibliography{biblio}


\appendix

\section{Exact equations relating the kernel and the auto-correlation functions}\label{sec:correl}
\label{sec:correl_exact}

To simplify notations, we write (in this appendix only) $\varphi(t) = K(t,t)$.

For $s<t$, one has 
$\mathcal{C}(t-s) = \lambda_\infty \overline \lambda  - {\overline \lambda}^2 + \EE{A_t \frac{\dd N_s}{\dd s}} + 2 \EE{M_t \frac{\dd N_s}{\dd s}}$.
\begin{equation}
\EE{A_t \frac{\dd N_s}{\dd s}} = \int_{-\infty}^t \varphi(t-u) \EE{\frac{\dd N_u}{\dd u} \frac{\dd N_s}{\dd s}} \dd u.
\nonumber
\end{equation}
For $u \neq s, \ \EE{\frac{\dd N_u}{\dd u} \frac{\dd N_s}{\dd s}} \dd u = [\mathcal{C}(u-s)+ {\overline \lambda}^2] \dd u$, and for 
$u=s, \ \EE{\left(\frac{\dd N_u}{\dd u}\right)^2} \dd u = \kappa \EE{\frac{\dd N_u}{(\dd u)^2}} \dd u = \kappa \overline{\lambda}$, where
$\kappa$ is the kurtosis of the law $\mu$ of the jumps of $P$ ($\kappa = 1$ if $\Delta P_\tau = \pm \psi$). Thus,
\begin{equation}
\EE{A_t \frac{\dd N_s}{\dd s}} = \text{Tr}(K) {\overline \lambda}^2 
+ \kappa \overline{\lambda} \varphi(t-s) + \int_{-\infty}^t \varphi(t-u) \mathcal{C}(u-s) \dd u.
\nonumber
\end{equation}
On the other hand,
\begin{align}
\EE{M_t \frac{\dd N_s}{\dd s}} &= \frac1{\psi^2} \int_{-\infty}^t \EE{\Theta_{t,u} \frac{\dd P_u}{\dd u} \frac{\dd N_s}{\dd s}} \dd u
\nonumber \\
&= \frac1{\psi^2} \int_{-\infty}^t \int_{-\infty}^{u-} K(t-u,t-r) \EE{\frac{\dd N_s}{\dd s} \frac{\dd P_u}{\dd u} \frac{\dd P_r}{\dd r}} \dd r \dd u
\nonumber \\
&= \int_{-\infty}^{s-} \int_{-\infty}^{u-} K(t-u,t-r) \mathcal{D}(s-u,s-r) \dd r \dd u,
\nonumber
\end{align}
since $\Delta P_\tau$ and $(\Delta P_\tau)^3$ are centered, which implies that
$\EE{\frac{\dd N_s}{\dd s} \frac{\dd P_u}{\dd u} \frac{\dd P_r}{\dd r}} = 0$ for $u \geq s$.
Taking $t = \tau>0$ and $s=0$, we obtain
\begin{equation}
\mathcal{C}(\tau) = \kappa \overline{\lambda} \varphi(\tau) + \int_{-\infty}^\tau \varphi(\tau-u) \mathcal{C}(u) \dd u
+ 2 \int_{0+}^\infty \int_{u+}^\infty K(\tau+u,\tau+r) \mathcal{D}(u,r) \dd r \dd u.
\nonumber
\end{equation}

For $t>t_1>t_2$, one has 
$\mathcal{D}(t-t_1,t-t_2) = \frac1{\psi^2} \EE{A_t \frac{\dd P_{t_1}}{\dd t_1} \frac{\dd P_{t_2}}{\dd t_2}} 
+ \frac2{\psi^2} \EE{M_t \frac{\dd P_{t_1}}{\dd t_1} \frac{\dd P_{t_2}}{\dd t_2}}$.
The first term gives
\begin{align}
\frac1{\psi^2} \EE{A_t \frac{\dd P_{t_1}}{\dd t_1} \frac{\dd P_{t_2}}{\dd t_2}} &= 
\frac1{\psi^2} \int_{-\infty}^t \varphi(t-u) \EE{\frac{\dd N_u}{\dd u} \frac{\dd P_{t_1}}{\dd t_1} \frac{\dd P_{t_2}}{\dd t_2}} \dd u
\nonumber \\
&= \int_{t_1+}^t \varphi(t-u) \mathcal{D}(u-t_1,u-t_2) \dd u.
\nonumber
\end{align}
The second term is given by
\begin{equation}
\frac1{\psi^2} \EE{M_t \frac{\dd P_{t_1}}{\dd t_1} \frac{\dd P_{t_2}}{\dd t_2}} = 
\frac1{\psi^4} \int_{-\infty}^t \int_{-\infty}^{u-} K(t-u,t-r) \EE{\frac{\dd P_{t_1}}{\dd t_1} \frac{\dd P_{t_2}}{\dd t_2} \frac{\dd P_u}{\dd u} \frac{\dd P_r}{\dd r}} \dd r \dd u
\nonumber
\end{equation}
Since $r<u$ in the integral and $t_2<t_1$, the expected value is zero if $u \neq t_1$. 
For $u=t_1$, we have 
$\EE{\left(\frac{\dd P_u}{\dd u}\right)^2 \frac{\dd P_{t_2}}{\dd t_2} \frac{\dd P_r}{\dd r}} \dd u
= \psi^2 \EE{\frac{\dd N_u}{(\dd u)^2} \frac{\dd P_{t_2}}{\dd t_2} \frac{\dd P_r}{\dd r}} \dd u
= \psi^2 \EE{\frac{\dd N_{t_1}}{\dd t_1} \frac{\dd P_{t_2}}{\dd t_2} \frac{\dd P_r}{\dd r}}$. Thus,
\begin{equation}
\EE{M_t \frac{\dd P_{t_1}}{\dd t_1} \frac{\dd P_{t_2}}{\dd t_2}} = 
\frac1{\psi^2} \int_{-\infty}^{t_1-} K(t-t_1,t-r) \EE{\frac{\dd N_{t_1}}{\dd t_1} \frac{\dd P_{t_2}}{\dd t_2} \frac{\dd P_r}{\dd r}} \dd r.
\nonumber
\end{equation}
For $r \neq t_2$, one has 
$\frac1{\psi^2} \EE{\frac{\dd N_{t_1}}{\dd t_1} \frac{\dd P_{t_2}}{\dd t_2} \frac{\dd P_r}{\dd r}} \dd r = \mathcal{D}(t_1-t_2,t_1-r) \dd r$.
On the other hand $r = t_2$ yields $\EE{\frac{\dd N_{t_1}}{\dd t_1} \frac{\dd N_r}{(\dd r)^2}} \dd r = 
\EE{\frac{\dd N_{t_1}}{\dd t_1} \frac{\dd N_{t_2}}{\dd t_2}} = \mathcal{C}(t_1-t_2)+{\overline{\lambda}}^2$. We obtain
\begin{equation}
\EE{M_t \frac{\dd P_{t_1}}{\dd t_1} \frac{\dd P_{t_2}}{\dd t_2}} =
K(t-t_1,t-t_2) [\mathcal{C}(t_1-t_2)+{\overline{\lambda}}^2]
+ \int_{-\infty}^{t_1-} K(t-t_1,t-r) \mathcal{D}(t_1-t_2,t_1-r) \dd r.
\nonumber
\end{equation}
We eventually obtain by taking $\tau_2 = t > \tau_1 = t-t_1, t_2 = 0$,
\begin{align}
\mathcal{D}(\tau_1,\tau_2) =
2 K(\tau_1,\tau_2) [\mathcal{C}(\tau_2-\tau_1)+{\overline{\lambda}}^2]
&+ \int_{(\tau_2-\tau_1)+}^{\tau_2} \varphi(\tau_2-u) \mathcal{D}(u-\tau_2+\tau_1,u) \dd u
\nonumber \\
&+ 2 \int_{-\infty}^{(\tau_2-\tau_1)-} K(\tau_1,\tau_2-u) \mathcal{D}(\tau_2-\tau_1,\tau_2-\tau_1-u) \dd u.
\nonumber
\end{align}

\section{Asymptotic analysis of the Hawkes + ZHawkes process}

In order to analyze the coupled Hawkes + ZHawkes processes, we first write the Fokker-Planck equation for the joint probability $\Pi(h,y)$ of 
$h \equiv \bar{H}$ and $y \equiv \bar{Y}=\bar{Z}^2$. Setting $t \leftarrow \beta t$ as the new time, we find:

\begin{align}
\label{FP}
\frac{\partial \Pi}{\partial t} 
&= - \frac{\partial}{\partial h} \left\{ \left[ -(1-n_H) h  + n_H (\lambda_\infty + y) \right] \Pi \right\}
\\ \nonumber 
&- \chi \frac{\partial}{\partial y} \left\{ \left[(n_Z - 1) y + n_Z (\lambda_\infty + h) \right] \Pi \right\}
+2 \chi n_Z \frac{\partial^2}{\partial y^2} \left\{ \left[ y (\lambda_\infty + h + y) \right] \Pi \right\}
\end{align}

We will study the stationary distribution of the process, such that the left-hand side of the above equation is zero. We introduce the conditional 
distribution of $h$ for a given $y$, $\Pi(h|y)$, and the marginal distribution of $y$, $\pi(y)$, as:
\be
\pi(y) := \int_0^\infty {\rm d}h \, \Pi(h,y); \qquad \Pi(h|y) = \frac{\Pi(h,y)}{\pi(y)},
\ee
and the generating function of $\Pi(h|y)$, as:
\be
Z(z|y)= \int_0^\infty {\rm d}h \, e^{-zh} \Pi(h|y),
\ee
such that $Z(0|y)=1$ and $Z'(0|y):=-a^*$ is the conditional average of $h$ for a given $y$. 

Now we assume, and self consistently check, that for large $y$, $\Pi(h|y)$ is of the form $1/y F(h/y)$, which means that $h$ is a random variable of order $y$.
This implies:
\be
Z(z|y) = G(x=zy); \qquad G(x):= \int_0^\infty {\rm d}u \, e^{-zu} \, F(u).
\ee

Multiplying Eq.(\ref{FP}) by $e^{-zh}$ and integrating over $h$ then leads, in the stationary state, to:

\begin{align}
&- x \pi(y) \left[ (1-n_H) G'(x)  + n_H G(x) \right] - \chi \frac{\partial}{\partial y} \left\{ \left[(n_Z - 1) G(x) - n_Z G'(x) \right] y \pi(y) \right\}
\\ \nonumber
&+ 2 \chi n_Z \frac{\partial^2}{\partial y^2} \left\{ \left[G(x) - G'(x) \right] y^2 \pi(y) \right\} = 0,
\end{align}
where we have assumed $y \gg \lambda_\infty$. In the asymptotic limit, $\pi(y)$ behaves as a power law: $\pi(y) \propto A/y^{1+\mu}$. Indeed, injecting this 
ansatz into the last equation leads to a non-trivial equation for $G(x)$ where $y$ and $A$ have completely disappeared:
\begin{align}\label{eq-G}
x \left[ (1-n_H) G'(x)  + n_H G(x) \right]  &=  \chi \left[ \mu n_Z H(x) - n_Z x H'(x) - \mu G(x) + x G'(x) \right] \\ \nonumber 
&+ 2 \chi n_Z \left[x^2 H''(x) + 2(1-\mu) x H'(x) -\mu(1-\mu) H(x) 
\right],
\end{align}
where we have introduced the shorthand $H(x)=G(x)-G'(x)$. Let us first analyze this equation for $x=0$; without any further assumptions one has, with $G(0)=1$ and $G'(0)=-a^*$:
\be
\mu n_Z (1+a^*) - \mu - 2n_Z \mu (1-\mu) (1+a^*) = 0 \Rightarrow \mu = \frac12 + \frac{1}{2 n_Z(1+a^*)},
\ee
where the unphysical solution $\mu=0$ was discarded. We thus need to solve Eq. (\ref{eq-G}) for $G(x)$ and determine $a^*$ from the value of $-G'(0)$. 
An easy case is $\chi=0$. One immediately finds: 
\be
(1-n_H) G'(x)  + n_H G(x) = 0 \Rightarrow G_0(x)= e^{-n_H x/(1-n_H)},
\ee
leading to $a_0^* = n_H/(1-n_H)$. The small $\chi$ expansion is also conveniently performed by setting $G(x) = G_0(x)+ \chi g_1(x) + \chi^2 g_2(x) + \dots$. To first order in $\chi$, 
the equation for $g_1$ reads:
\be
(1-n_H) g_1'(x)  + n_H g_1(x)  = G_0(x)\left[ \frac{n_H(1-n_H-n_Z)}{(1-n_H)^2} + \frac{n_H^2}{(1-n_H)^2} x \right],
\ee
and thus, with the right boundary condition for $g_1(x)$, 
\be
g_1(x) = \left[ \frac{n_H(1-n_H-n_Z)}{(1-n_H)^3} x + \frac{n_H^2}{2(1-n_H)^3} x^2 \right] e^{-a_0^* x}.
\ee
To first order, one thus finds:
\be
a^* = \frac{n_H}{(1-n_H)} \left[ 1 - \chi \frac{(1-n_H-n_Z)}{(1-n_H)^2}  + O(\chi^2) \right]. 
\ee

In the opposite limit $\chi \to \infty$, one finds that $G(x)=1$ solves the equation, as expected since in this limit $h$ cannot follow the dynamics of $y$, and therefore
one expects that in the limit $y \to \infty$, $F(u) \approx \delta(u)$ and thus $G(x)=1$. When $\chi$ is large but not infinite, one can expect that $F(u)$ has a width of
order $\chi^{-1}$, and thus that $G(x)$ is a function of $x/\chi$. This means that each derivative of $G$ brings an extra factor $\chi^{-1}$. Setting $a^* = a/\chi$ and matching 
the terms in Eq. (\ref{eq-G}), we find:
\be
\chi (\mu + x (1-n_Z) ) G'(x) + (a \mu + n_H x) G(x),
\ee
or:
\be
\ln G(x)= - \frac{1}{\chi (1-n_Z)} \left[ n_H x + \mu ( a - n_H/(1-n_Z) ) \ln ( \mu + (1-n_Z) x ) \right],
\ee
which shows that our assumption that $G(x)$ is a function of $x/\chi$ singles out $a = n_H/(1-n_Z)$ as the only possibility, in which case:
\be
G(x) =_{\chi \to \infty} e^{-\frac{n_H x}{\chi (1-n_Z)}}.
\ee
This means that in this limit, $\Pi(h|y) \approx \delta(h - \frac{n_H}{\chi (1-n_Z)} y)$.

Finally, let us consider the limit $n_Z \to 0$ for a finite $\chi$. The idea now is to postulate that for small $n_Z$, $\Pi(h|y)$ is stronlgy peaked around $a^* y$, 
with a width that goes to zero as $\sqrt{n_Z}$. This translates into the following ansatz for $G(x)$
\be
G(x) = e^{-a^* x} {\cal G}(\sqrt{n_Z} x).
\ee
We can now analyze Eq. (\ref{eq-G}) in the regime $n_Z \to 0$ with fixed $z = \sqrt{n_Z} x$. The leading order terms are of order $1/n_Z$, and 
lead to an equation that is identically satisfied. The next two orders, $O(1/\sqrt{n_Z})$ and $O(1)$ allow us to fix both the function ${\cal G}(z)$ and the value of $a^*$.
We find in particular:
\be
{\cal G}(z) = \exp \left[\frac{(1+a^*)^2 [(1 - n_H + \chi) a^* - n_H]}{\chi} z^2 \right],
\ee
which shows that the distribution $\Pi(h|y)$ is in fact gaussian in that limit. We also find that $a^*$ obeys the following equation:
\be
( \gamma a^* - n_H) ( \gamma + (\gamma + 2 \chi) a^*) = a^{*2} \chi^2 \qquad \gamma = 1 - n_H + \chi.
\ee
The solution takes a simple form in the limits $\chi \to 0$ and $\chi \to \infty$, where we recover the results obtained above. 

In the general case, Eq. (\ref{eq-G}) is a third order, linear ODE for $G(x)$; imposing the correct boundary condition $G(x \to \infty)$ selects special values of $a^*$
for any triplet $(n_H,n_Z,\chi)$. The largest admissible value of $a^*$ corresponding to the smallest value of the tail exponent will be the physical solution. Unfortunately,
we have not been able to make progress yet on this general case.

 \end{document}